\documentclass[manuscript, screen]{acmart}
\usepackage{comment}

\AtBeginDocument{%
  \providecommand\BibTeX{{%
    \normalfont B\kern-0.5em{\scshape i\kern-0.25em b}\kern-0.8em\TeX}}}

\usepackage{xspace}
\newcommand{\system}{Sunnie\xspace}

\usepackage[shortlabels]{enumitem}
\usepackage{arydshln}
\usepackage{subfig}

\begin{abstract}
The human-computer interaction (HCI) research community has a longstanding interest in exploring the mismatch between users' actual experiences and expectation toward new technologies, for instance, large language models (LLMs). 
In this study, we compared users’ (N = 38) initial expectations against their post-interaction perceptions of two LLM-powered mental well-being intervention activity recommendation systems. Both systems have a build-in LLM to recommend personalized well-being intervention activity, but one system (\system) has an anthropomorphic conversational interaction design via elements such as appearance, persona, and natural conversation. Results showed that user engagement was high with both systems, and both systems exceeded users’ expectations along the utility dimension, highlighting AI’s potential to offer useful intervention activity recommendations. 
In addition, \system further outperformed the non-anthropomorphic baseline system in relational warmth. 
These findings suggest that anthropomorphic conversational interaction design may be particularly effective in fostering warmth in mental health support contexts. 

\end{abstract}

\setcopyright{acmlicensed}
\copyrightyear{2024}
\acmYear{2024}

\acmConference[Conference acronym 'XX]{Make sure to enter the correct
  conference title from your rights confirmation email}{June 03--05,
  2018}{Woodstock, NY}
\acmISBN{978-1-4503-XXXX-X/18/06}

\author{Siyi Wu}
\authornote{Both authors contributed equally to the paper}
\affiliation{
    \institution{University of Toronto}
    \city{Toronto}
    \country{Canada}}
\email{reyna.wu@mail.utoronto.ca}

\author{Julie Y. A. Cachia}
\authornotemark[1]
\affiliation{
    \institution{Flourish Science}
    \city{California}
    \country{USA}}
\email{julie@flouriship.com}

\author{Feixue Han}
\affiliation{
    \institution{Northeastern University}
    \city{Boston}
    \country{USA}}
\email{han.feix@northeastern.edu}

\author{Bingsheng Yao}
\affiliation{
    \institution{Northeastern University}
    \city{Boston}
    \country{USA}}
\email{b.yao@northeastern.edu}

\author{Tianyi Xie}
\affiliation{
    \institution{Flourish Science}
    \city{California}
    \country{USA}}
\email{t@flouriship.com}

\author{Xuan Zhao}
\affiliation{
    \institution{Flourish Science}
    \city{California}
    \country{USA}}
\email{xuan@flouriship.com}

\author{Dakuo Wang}
\affiliation{
    \institution{Northeastern University}
    \city{Boston}
    \country{USA}}
\email{d.wang@northeastern.edu}

\renewcommand{\shortauthors}{Wu, et al.}



\begin{teaserfigure}
  \includegraphics[width=\textwidth]{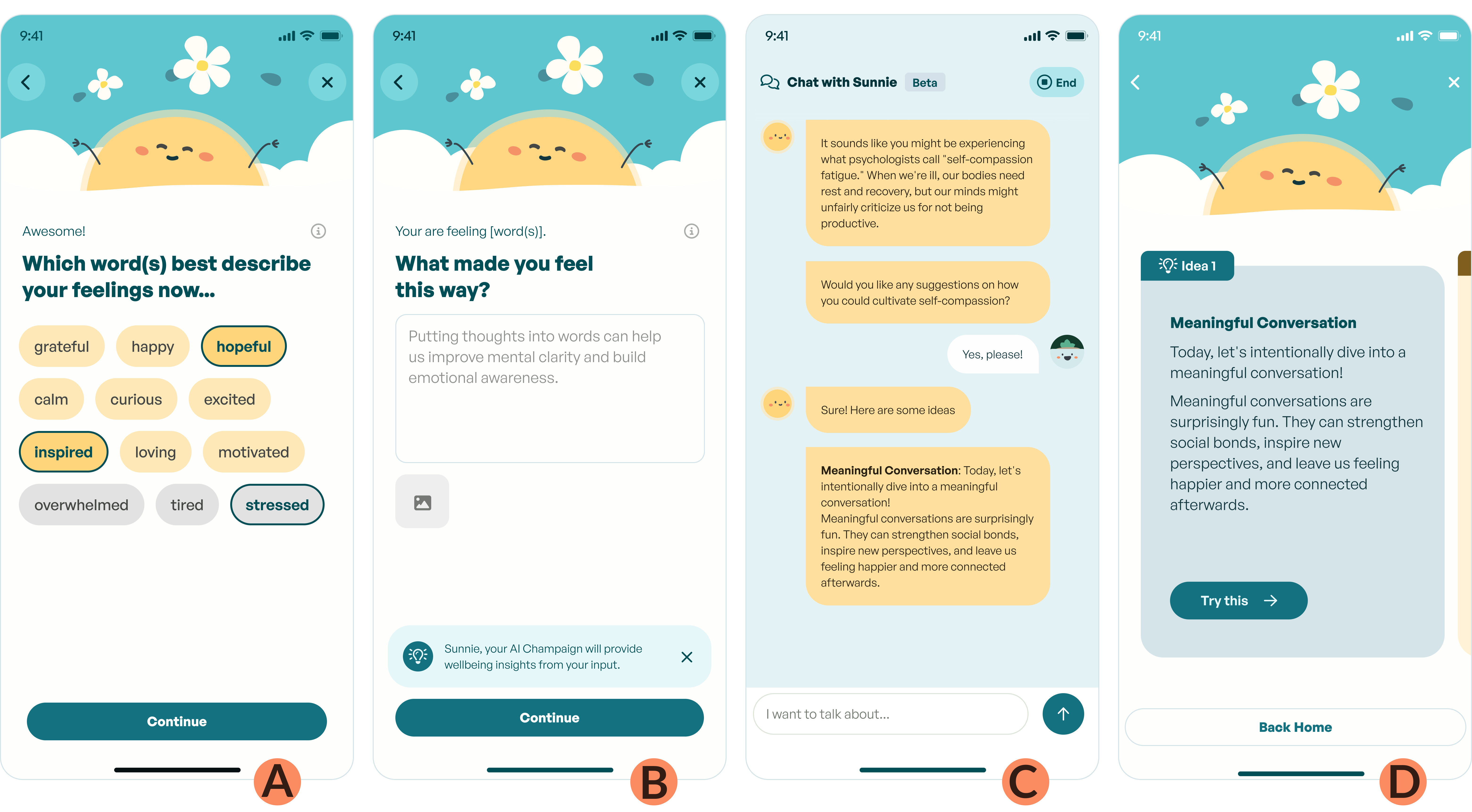}
  \caption{\system is an anthropomorphic LLM-based conversational agent, designed to improve well-being through personalized micro-activity recommendations. 
  \system consists of three key features: the anthropomorphic design of \system's appearance and the persona prompts, the multi-turn natural conversation of well-being coaching, and the LLM-based personalized activity recommendation module. 
  The four panels depict users' interactions with \system: (A) users select the best word(s) that describe their feelings, (B) users enter a description of the perceived source of their feelings, (C) users interact with \system in a multi-turn conversation, and (D) \system recommends a personalized activity to support users' well-being. }
  \label{fig:teaser}
  \Description{The teaser figure shows four user interfaces of Sunnie, which is the proposed anthropomorphic LLM-based conversational agent, on a mobile app screen. Four user interfaces are marked with ``A'', ``B'', ``C'', and ``D'', respectively. 
  Screen A shows the interface that asks users ``Which word(s) best describe your feelings now...'', with 13 pre-defined options visualized as clickable buttons. Nine out of 13 buttons are positive feelings (grateful, happy, hopeful, calm, curious, excited, inspired, loving, and motivated) which are of yellow background color, whereas the other four buttons are negative feelings (overwhelmed, tired, stressed, others), which are of grey background color. Users are allowed to select multiple buttons. There is a Sunnie's appearance at the top of the interface and a ``continue'' button at the bottom.
  Screen B shows the interface that asks users ``What made you feel this way'', and provides a text input box for users' additional input. There is a Sunnie's appearance at the top of the interface and a ``continue'' button at the bottom.
  Screen C shows the interface for chatting with Sunnie. It is a chatbox interface that allows users to establish multi-turn communications with Sunnie. On the screen, Sunnie first asks ``It sounds like you might be experiencing what psychologists call `self-compassion fatigue.' When we are sick, our bodies need rest and recovery, but our minds might unfairly criticize us for not being productive.'', and ``Would you like any suggestions on how you could cultivate self-compassion?''. The user replies ``Yes, please!''. Sunnie continues the conversation with ``Sure! Here are some ideas.'' and ``Meaningful Conversation: Today, let's intentionally dive into a meaningful conversation! Meaningful conversations are surprisingly fun. They can strengthen social bonds, inspire new perspectives, and leave us feeling happier and more connected afterwards.''
  Screen D shows the interface for activity recommendations. The recommended activity is `Meaningful Conversation'' with a short description: ``Today, let's intentionally dive into a meaningful conversation! Meaningful conversations are surprisingly fun. They can strengthen social bonds, inspire new perspectives, and leave us feeling happier and more connected afterwards.''. There is a Sunnie's appearance at the top of the interface a ``Try this'' button below the activity description, and a ``Back Home'' button at the bottom.}
\end{teaserfigure}

\begin{document}

\title[``I Like \system More Than I Expected!'']{``I Like \system More Than I Expected!'':  Exploring User Expectation and Perception of an Anthropomorphic LLM-based Conversational Agent for Well-Being Support}






\maketitle

\section{Introduction}

In recent years, mental well-being support has become increasingly critical as demand for services continues to grow while access to mental health professionals remains limited \cite{who2023, apa2022impact}. In response, a variety of evidence-based intervention activities have been developed~\cite{lyubomirsky2011becoming, sheldon2006increase}. These are small, manageable actions designed to improve mental well-being, based on established frameworks such as the PERMA model \cite{seligman2011} and other research on physical and cognitive activities \cite{walsh2011lifestyle, Penedo2005ExerciseAW}. However, despite the wealth of information on activities to improve well-being, most people still find it difficult to integrate these activities into their daily routines, even when they understand the potential benefits~\cite{Schueller2013, Prochaska1997}. 
This gap between awareness and action (i.e., the knowledge-action gap) can be attributed to several factors, including lack of motivation, perceived difficulty in adopting new behaviors, and the absence of personalized guidance~\cite{knittle2018can, Ryan2000, Prochaska1997}. 

One promising approach to bridge the knowledge-action gap is through mental well-being activity recommendation systems, which use algorithms to promote positive behaviors and well-being~\cite{li2023systematic, tosti2024using, inkster2018empathy, abd2020effectiveness, grove2021co, zhang2020artificial, liu2022using}. Approaches such as gamified apps and music-based technologies have been explored for their ability to create interactive experiences that reinforce prosocial behaviors~\cite{walsh2011lifestyle, Penedo2005ExerciseAW}. However, these systems often rely on passive interactions, focusing on external rewards or mood modulation without fostering deep emotional engagement or two-way communication~\cite{kwak2024investigating, turan2016older, sanchez2014acceptability, boduszek2019prosocial}. To create more interactive and engaging experiences, conversational agents have emerged as a solution, offering conversational interactions that encourage user engagement and support the adoption of healthier behaviors ~\cite{Oh2021ASR, Aggarwal_Tam_Wu_Li_Qiao_2023, stephens2019feasibility, piao2020use, kramer2020components, kramer2019investigating, kunzler2019exploring, kocielnik2018reflection, zhang2020artificial, maher2020physical, fadhil2019assistive, Oh2021ASR, Krame_2020}. Research has shown their utility in improving lifestyle habits, such as promoting physical activity, better diet, and improved sleep quality ~\cite{kramer2020components, Aggarwal_Tam_Wu_Li_Qiao_2023, Oh2021ASR, Singh2023SystematicRA, Perski2019DoesTA, fan2021utilization, wang2021cass}. Despite these advances, conversational agents still face challenges in delivering highly personalized and contextually relevant suggestions, often requiring significant user input and struggling to adapt to evolving user needs~\cite{kramer2020components, Aggarwal_Tam_Wu_Li_Qiao_2023, Oh2021ASR, Singh2023SystematicRA, perski2017conceptualising, zhang2022storybuddy, Schueller2013, Prochaska1997}.

The emergence of large language models (LLMs) presents new possibilities to enhance mental well-being intervention activities recommendation systems, with their superior natural language interpretation capabilities, which could foster a more nuanced understanding of context, recommend appropriate intervention activities, and provide interactions through conversations~\cite{de2023benefits, torous2024generative, van2023global, yang2024talk2care, mahmood2023llm, chan2023mango}.
Such advantages of LLMs could lead to more acceptable intervention activity recommendation and more human-like user experiences, potentially increasing user engagement and adherence\cite{loh2023harnessing, ma2023understanding, Song2024TheTC}. 
In addition, the integration of anthropomorphic design with conversational interfaces, such as avatars, facial expressions, and personalized conversational styles, has the potential to create interactions that feel more natural and emotionally resonant~\cite{seeger2018designing, laban2021perceptions, Bowman_Cooney_Newbold_Thieme_Clark_Doherty_Cowan_2024, chinmulgund2023anthropomorphism}. Despite the potential of LLM-based recommendation systems, there is limited research on their use in promoting intervention activities for well-being support. 
Most existing studies focus on improving personalization and providing a more human-like experience through conversations but rarely focus on facilitating action-taking or recommending activities to practice well-being~\cite{Song2024TheTC, Hua2024LargeLM, ma2023understanding, loh2023harnessing, liu2023chatcounselor, yao2023development, cho2023evaluating}. This combination of LLM-based recommendation and anthropomorphic interaction offers a unique opportunity for mental well-being activity recommendation systems that make personalized recommendations and foster meaningful and engaging interactions, addressing both personalization and user experience gaps in current systems.

Furthermore, there has been substantial research interest in the HCI community to explore the discrepancy between users' actual experience and expectations toward a novel technology, and how HCI researchers can bridge the gap with various design strategies.  
Previous research has shown that users often hold expectations that AI systems are high in agency (e.g., capacity for self-control) but low in experience (e.g., capacity to feel)~\cite{gray2007dimensions}. These expectations might limit how much users are willing to rely on AI-driven mental health support, which may be perceived to stem from the capacity to feel. While previous research has examined expectations versus experience in other domains such as gaming ~\cite{michalco2015relation} and conversational agents in various contexts~\cite{luger2016like, cho2019once}, we build on this line of work to examine users' experience and expectations within the context of well-being applications and to specifically compare anthropomorphic versus non-anthropomorphic designs.

Thus, a critical yet unaddressed question is how: 1) the design of LLM-based mental well-being intervention activity recommendation systems, especially the incorporation of anthropomorphic features, influences users' perceptions and their willingness to engage in actions that support well-being, and 2) whether and to what extent users underestimate the capabilities of these systems.

Thus, our study is guided by three research questions:
\begin{itemize}
    \item \textbf{RQ1}: How do anthropomorphic designs of LLM-based mental well-being intervention activity recommendation systems shape users' \textbf{perceptions} of such systems?
    \item \textbf{RQ2}: How do such designs influence users' \textbf{engagement} in adopting recommended mental well-being activities? 
    \item \textbf{RQ3}: To what extent do users' \textbf{expectations} of LLM-based systems align with their actual experiences of these systems?
\end{itemize}

We introduce \system, an anthropomorphic LLM-based conversational agent designed to offer personalized well-being support and recommend practical actions grounded in positive psychology and social psychology research. 
\system is an ``AI happiness coach and companion'' developed by Flourish Science, a public benefit corporation with the mission of ``personalizing the science of happiness and well-being to make it accessible, actionable, simple, and fun, thereby providing proactive and just-in-time mental health and well-being support.'' The version tested in the current study is powered by GPT-4~\cite{achiam2023gpt}, an advanced LLM, to offer more human-like interactions, thereby enhancing personalization and encouraging users to take action to improve their well-being.

We hypothesize that incorporating anthropomorphic features into the design of Sunnie will lead to more positive perceptions, which, in turn, will increase user engagement and the likelihood of adopting the recommended well-being activities. Additionally, we hypothesize that users may underestimate the capabilities of the anthropomorphic design, particularly in its ability to convey warmth. To test these hypotheses, we conducted an empirical user study that compared participants' expectations and perceptions of two different systems: Sunnie, an anthropomorphic LLM-based chatbot, and a non-anthropomorphic, non-conversational, LLM-based activity recommendation system.
We used proxy metrics based on participants' perceptions of the system and their self-reported engagement with recommended activities. This approach aligns with prior research, which suggests that one-time interactions can provide valuable insights into potential long-term adoption and behavior change~\cite{yardley2016understanding, bijkerk2023measuring, consolvo2009theory}.

The core contributions of this paper are tri-fold:
\begin{itemize}
    \item Describe the design and development of \system, an anthropomorphic LLM-based conversational agent equipped with emotion regulation coaching and well-being activity recommendations, aimed at enhancing personalized well-being support and facilitating proactive well-being management.
    \item Report findings and insights from an empirical 3-day, within-subject user study through quantitative and qualitative analysis examining how anthropomorphic designs shape user perceptions, engagement, and how these perceptions align with their initial expectations. 
    \item Provide design considerations derived from our study results to foster the future development of LLM-based well-being support systems in terms of more personalized and effective support.
    
\end{itemize}

\section{Related Work}
\subsection{AI in Activity Recommendation for Well-Being Support}
While abundant information is available on activities that promote physical and psychological well-being, a significant gap exists in adherence and motivation to participate in these activities. AI-powered technologies, particularly chatbots, have been widely used to address this gap by promoting activities for well-being support~\cite{stephens2019feasibility, piao2020use, kramer2020components, kramer2019investigating, kunzler2019exploring, kocielnik2018reflection, zhang2020artificial, maher2020physical, fadhil2019assistive, Oh2021ASR, Krame_2020}.
Building on this foundation, most studies have focused on encouraging physical activities, demonstrating the effectiveness of using chatbots to guide people toward a healthy lifestyle, such as promoting healthy diets, improving the duration and quality of sleep, quitting smoking, etc.~\cite{kramer2020components, Aggarwal_Tam_Wu_Li_Qiao_2023, Oh2021ASR, Singh2023SystematicRA, Perski2019DoesTA}. These interventions have proven to be effective in various populations and age groups across both short-term and long-term studies. 

Expanding the scope, while research has shown the importance of activities promoting positive emotions and behaviors for well-being~\cite{Ly_Ly_Andersson_2017}, a few studies focused on using chatbots to deliver positive psychology and Cognitive Behavior Therapy (CBT) interventions. For example, Kien et al. \cite{Ly_Ly_Andersson_2017} studied the effectiveness and adherence of using chatbots to deliver positive psychology and CBT strategies; Rohani et al. \cite{Rohani2020MUBSAP} showed the positive impacts of MUBS \cite{Rohani2020MUBSAP}, a smartphone-based system supporting Behavioral Activation (BA) treatment of depressive symptoms with a personalized content-based activity recommendation model based on multinomial Naive Bayes machine learning algorithms, in motivating patients to engage in pleasant activities. 

However, despite the potential of chatbots in promoting well-being activities, the challenges remain to ensure their effectiveness in user adherence. These challenges include the need for improved linguistic capabilities, more personalized content, and the integration of human-like identity features to improve user experience and engagement~\cite{Abd-Alrazaq2020, laranjo2018conversational, vaidyam2019chatbots}. 

Building on these gaps, our work aims to contribute to the field by studying the effectiveness of well-being activities recommended by a conversational agent with more human-like features, personalized recommendations, and improved linguistic capabilities using LLMs coupled with anthropomorphic designs. We seek to explore how these features can enhance the effectiveness and adherence of chatbot interventions for well-being support, addressing the current limitations in chatbot personalization and user engagement.

\subsection{LLM-Based Systems for Well-Being Support}

The use of conversational agents to enhance well-being has a longstanding history ~\cite{Bowman_Cooney_Newbold_Thieme_Clark_Doherty_Cowan_2024,Abd-Alrazaq2020}, dating back to pioneering systems such as ELIZA~\cite{weizenbaum1966eliza}. This tradition continues with the advent of modern chatbots such as WoeBot~\cite{Fitzpatrick2017} and Wysa~\cite{Inkster2018AnEC}, which are easily accessible to the public. 

In the past, mental health chatbots primarily utilized rule-based systems~\cite{Abd-Alrazaq2020}, employing various therapeutic techniques to guide users through self-help exercises. These chatbots have been shown effective in enhancing well-being by encouraging self-disclosure \cite{Lee20chi, Lee20}, fostering self-compassion~\cite{Minha19}, and regulating users' emotions~\cite{Denecke21}, etc. However, the rule-based nature of these chatbots often limits the natural flow of conversation~\cite{Song2024TheTC}. 

The introduction of LLMs has sparked a new wave of interest in the potential of LLM-based conversational agents for mental health support~\cite{Song2024TheTC}, such as platforms like OpenAI's ChatGPT~\cite{achiam2023gpt} and Replika~\cite{laestadius2022too}. The user-friendly conversational interfaces of these LLM-powered chatbots have sparked excitement among clinicians about the possibilities of novel AI-driven interventions \cite{Song2024TheTC}. These agents are designed to provide direct interaction with individuals seeking mental health support through various platforms, including personal digital companions \cite{ma2023understanding}, online on-demand counseling \cite{loh2023harnessing, liu2023chatcounselor, yao2023development, liu2023taskadaptive, lee2023chain, cho2023evaluating, zhang-etal-2023-ask}, emotional support~\cite{zheng2023building}, etc.





Building on this foundation, our work aims to expand the utility of LLM-based conversational agents by incorporating activity recommendations alongside conversational support. By integrating more human-like features, personalized recommendations, and enhanced linguistic capabilities, we seek to explore how these features can enhance the effectiveness and adherence of chatbot interventions for well-being support, addressing the current limitations in chatbot personalization and user engagement.




\subsection{Anthropomorphism Design of LLM-Based Conversational Agent}

One of the limitations of LLM-based conversational agents in mental health is concern about trust and safety~\cite{Abd-Alrazaq2020}. 
Trust is a fundamental element in mental health support, and ensuring the safety and reliability of conversational agents is crucial for their acceptance and effectiveness. This concern highlights the need for careful design and ethical considerations in their development \cite{bickmore2001relational, Lee20, li2023influence}. Anthropomorphism refers to the psychological phenomenon of ``attributing human characteristics to the
nonhuman''~\cite{seeger2018designing}, is one aspect of this need that should be used with care, as it influences user expectations and reliance on AI systems, affecting how users perceive and interact with conversational agents \cite{ma2023understanding}. 



In the anthropomorphic design for conversational agents, features can be broadly categorized into social and verbal cues~\cite{seeger2018designing, Pradhan2021, laban2021perceptions, Bowman_Cooney_Newbold_Thieme_Clark_Doherty_Cowan_2024, chinmulgund2023anthropomorphism}. Social cues encompass non-verbal elements that convey human-like traits and behaviors, such as human-like appearance, including facial expressions and gestures, interactivity that mimics human responsiveness, and behavioral features that reflect human personality, empathy, and social roles. These cues enhance the perceived humanness of the agent, making interactions more relatable and engaging~\cite{li2023influence, bickmore2001relational, bickmore2005s}. On the other hand, verbal cues involve the use of language and communication styles that emulate human-like speech and interaction. This includes using natural language for intuitive and relatable communication, adherence to social norms such as politeness~\cite{Bowman_Cooney_Newbold_Thieme_Clark_Doherty_Cowan_2024}, greetings, and farewells, and providing advice in a manner consistent with human conversational patterns~\cite{clark2019makes, bickmore2010making, li2023influence}. 

The appropriate anthropomorphic design can amplify social responses and build social relationships between humans and computers. By incorporating human nature and unique traits, as well as personality traits, the perceived human likeness of systems is increased, which can improve user engagement and satisfaction~\cite{seeger2018designing, strohmann2023toward}. However, the implementation of anthropomorphic design must be balanced to avoid the uncanny valley phenomenon~\cite{wang2015uncanny}, where overly human-like design features can elicit feelings of eeriness or discomfort. This highlights the need for a nuanced approach to human-like design in conversational agents to ensure positive user perceptions and acceptance.

Research has explored how the design of AI systems influences people's perceptions in various settings, including clinical, social support, and public health interventions. However, there is a notable gap in understanding the specific influence of anthropomorphic designs on conversational agents, especially those aimed at fostering well-being activities. The significance of grasping how users perceive these agents is critical for the development of AI interfaces that are not only effective, but also provide a sense of care and support. In addressing this gap, our research delves into the effects of anthropomorphic design elements on the user's perception of an AI-powered companion dedicated to promoting well-being practices.


To advance this area of study, we first introduced the design of \system, an LLM-powered conversational agent that offers personalized guidance on emotion regulation and recommends relevant activities grounded in psychological research to improve well-being. To this end, we meticulously described its system architecture, design principles, user interface, and prompting framework. Next, we conducted an evaluation to understand people's perception of and experience with this conversational agent. Our approach stands out in its comprehensive consideration of how the integration of human-like characteristics within AI can transform the user experience, leading to a more positive and engaging interaction with technology aimed at supporting well-being and personal growth.

\begin{figure*}[!tp]
    \centering
    \includegraphics[width=.9\textwidth]{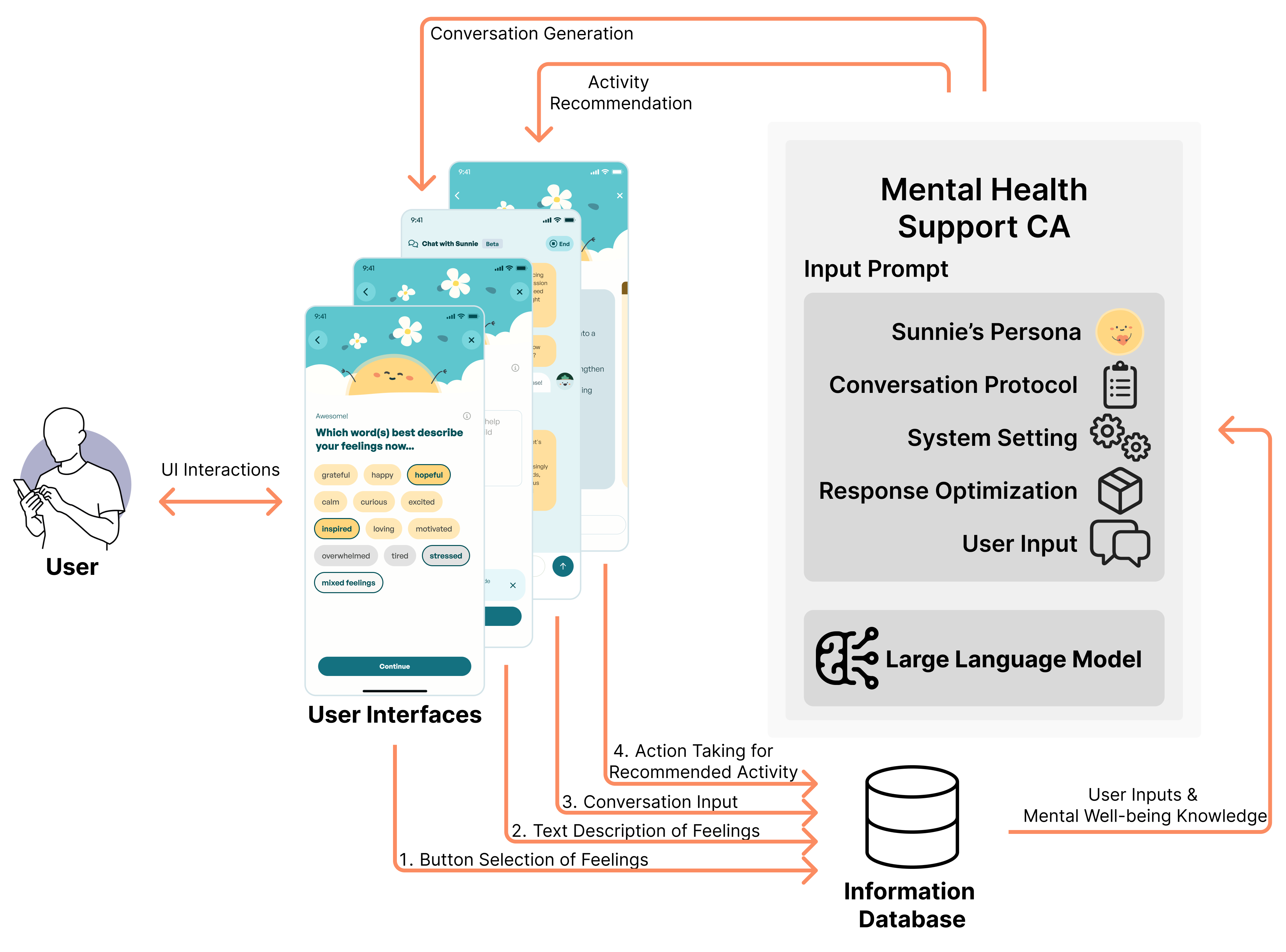}
    \caption{The system architecture of Sunnie. Sunnie is supported by an LLM with a meticulously designed prompt framework. A user interacts with Sunnie through a series of interfaces, including selecting and typing their feelings with buttons and text input, communicating with Sunnie in multi-turn conversation, receiving personalized activity recommendations, and deciding whether to take the activity.}
    \label{fig:system-architecture}
    \Description{Figure 2 shows the system architecture of Sunnie. The system works in the following flow: 1) A user interacts with Sunnie through a series of user interfaces, 2) the user interaction data are then fed into a standalone information database, 3) the information database sends the user inputs and well-being knowledge into the input prompt framework, 4) the input prompt framework constructs the complete input to a large language model (LLM), and the LLM generates responses that will be visualized back on the user interface. 
    There are four primary steps in the user interactions: 1) button selection of feelings, 2) text description of feelings, 3) conversation input, and 4) action taking for recommended activity. 
    There are two types of LLM outputs: 1) after the user's conversation input, the LLM generates conversation responses; and 2) after the user finishes the conversation with Sunnie, the LLM will generate a personalized activity recommendation to the user.}
\end{figure*}

\section{\system: An Anthropomorphic LLM-Based Conversational Agent }

\system aims to leverage the potential of anthropomorphic design in LLM-based conversational agents to recommend activities that support well-being. 
This approach goes beyond the current state-of-the-art (SOTA) methods by not merely offering activity suggestions via chatbots but by capitalizing on the sophisticated capabilities of LLM-based agents to forge a deeply personalized and engaging interaction with users.


In the following sections, we delve into the design and developmental framework of \system. 
We systematically unpack the architecture of \system, beginning with an overview of the foundational design principles that guide this system. 
Following this, we highlight the distinctive features that set \system apart, providing a detailed exploration of the innovative aspects that enhance user engagement and personalization.
Finally, we articulate the methodological approach employed in the implementation of \system, ensuring a coherent and robust application of these principles and features in practice.

\subsection{Design Principles}

The ``Computers are Social Actors'' (CASA) paradigm~\cite{nass1994computers, nass2000machines}, also known as ``social response theory,'' suggests that users respond with social behavior and attributions when machines exhibit human-like features such as interactivity, natural language use, or human-like appearance. 
Initially proposed by Nass and Moon~\cite{nass2000machines}, this paradigm extends to conversational agents (CAs)~\cite{strohmann2023toward, porcheron2018voice, reeves2018not, bluvstein2024imperfectly}
, with research indicating that human-like behavior and social cues \cite{feine2019taxonomy, seeger2018designing} 
in CAs can enhance social reactions, establish trust, and lead to perceptions of reliability. 
Critically, the literature reveals a nuanced understanding of how and why these phenomena occur.
Firstly, it has been found that the more a CA resembles a human in the interactions, the more natural and effortless the user's response tends to be ~\cite{bickmore2005establishing, de2016almost, strohmann2023toward}. 
The naturalness of interaction is important for establishing trust between humans and CAs.
Secondly, CAs' social cues have been shown to be effective in facilitating the development of trust and influencing the user's perspective of CAs, which is pivotal when humans rely on CAs' for decision-making~\cite{bickmore2005establishing, de2016almost, strohmann2023toward}. 
Additionally, many aspects of human-to-human relationships are transferable to the relationship between humans and CAs, such as establishing parasocial relationships where only one party extends emotional energy, interest, and time. Still, the other side is completely unaware of the other's effort~\cite{stever2017parasocial, mctear2018conversational, strohmann2023toward}. 

\textbf{DP1: Anthropomorphic designs: no more, no less}. 
Building on these insights, recent research has highlighted the importance of anthropomorphic design in conversational agents (CAs), taking into account the uncanny valley effect~\cite{gnewuch2018faster, yuan2019crossing, mori1970uncanny,seeger2018designing}, which suggests that overly human-like agents can elicit discomfort. 
A balanced approach to anthropomorphic design is recommended, considering human identity (such as human-like representation, gender, or age), non-verbal features (such as hand gestures, facial expressions, or emojis), and verbal characteristics (such as word choice and sentence structure)~\cite{seeger2018designing, strohmann2023toward}. 
Studies have shown that incorporating human-like visual representations, verbal cues like self-references or emotional expressions, and non-verbal behavior like emoticons or turn-taking can enhance the perception of human-likeness in CAs~\cite{seeger2018designing, li2023influence}. 
For example, enabling agents to use social dialogue, express emotions, and refer to themselves as ``I'' can make them appear more human-like. Non-verbal cues, such as blinking dots to communicate thinking gestures or emoticons to convey emotional expressions, also significantly create a human-like impression~\cite{li2023influence, seeger2018designing}. 

Another common practice with the goal of imbuing human-like traits in CAs is to create personas for the agent. A persona could be a fictional character with a name, age, or even a defined backstory and personality~\cite{Pradhan2021}. 
Some research argues that having a distinct persona or personality could contribute to a cohesive and consistent presence of the conversational agent for users and increase trust and the intention to use the technology~\cite{ho2018psychological, isbister2000consistency}. However, careful designs of personas are needed~\cite{Pradhan2021}. 
It is important to avoid reinforcing stereotypes or biases, ensuring that the personas are diverse and inclusive~\cite{Pradhan2021,strohmann2023toward}.  

Studies indicate that a ``more is more'' approach is not advisable, as it can negatively affect perceived anthropomorphism. 
For example, a regression analysis~\cite{seeger2018designing} revealed two significant interactions: one between non-verbal and verbal cues and another between non-verbal and human identity cues. 
These findings suggest that the combination of these design dimensions provides a consistent representation of human-likeness, but including all three dimensions do not necessarily increase users' perceptions of anthropomorphism~\cite{seeger2018designing}.


\textbf{DP2: Grounding conversation in science but with layman's languages.} 
In recent years, the study of happiness and well-being has witnessed significant growth from cognitive and psychological perspectives. This body of literature demonstrates that individuals can intentionally cultivate well-being through specific practices~\cite{VanderWeele2017, VanderWeele2019, Keyes2002}. 
However, the vast wealth of knowledge emerging from this line of research is not easily accessible to the general public, presenting a notable barrier to its application in everyday life.
We envision an immense potential in leveraging LLMs to enhance the accessibility of scientific insights through LLM's recommendation and conversation capacities, making the psychological insights from positive psychology and behavioral science accessible and actionable to a broader audience.


Additionally, the burgeoning field of research in CAs has underscored the importance of crafting dialogues based on mutual understanding and empirical evidence~\cite{strohmann2023toward, park2012law}. 
Using familiar language and clear, simple expressions can enhance comprehension and engagement, as suggested by previous literature~\cite{gnewuch2017towards, kreuter1999understanding}. The concept of mental health literacy further emphasizes the importance of using language that aids in the recognition and management of mental health issues~\cite{jorm1997mental}. 
These considerations underscore the vital relationship between language, understanding, and action in well-being support.

Building upon this foundation, conversational agents that support evidence-based communication become essential. Ensuring that the information provided is relevant, practical, and scientifically accurate is critical for enhancing the credibility of the conversational agent. Grounding conversations in science and using familiar language aligns with the principles of evidence-based practice, which advocate for integrating research evidence into decision-making processes. This design consideration could enhance the user experience and ensure that users receive validated and reliable information, thereby promoting well-being effectively.

\textbf{DP3: Considering positive design principles that align with system goals. }
To align with our objectives of enhancing well-being, we integrate positive design principles \cite{zhang2007toward}.
These principles are predicated on the notion that each design aspect should be oriented toward enriching the user experience. 
We aim to bolster well-being and happiness, ensuring the design harmonizes with our system's overarching goals. We follow various strategies encompassed in the aforementioned design principles, such as Design for Pleasure, Design for Personal Significance, and Design for Virtue.
We chose these design principles to ensure that our system design meets functional requirements and contributes positively to users' psychological state, fostering better well-being.

\subsection{Key Features}
In this section, we delve into the key features and anthropomorphic designs of \system as shown in Figure.~\ref{fig:teaser}, drawing on a comprehensive review of existing literature and foundational design principles. 

The key features include 1) anthropomorphic designs of \system, 2) an LLM-based conversational agent for well-being coaching, and 3) LLM-driven, personalized well-being activities recommendation. 

In accordance with DP1, \system's anthropomorphic designs are crafted to endow the system with human-like characteristics without crossing into the uncanny valley. These design elements aim to make interactions more natural and engaging by providing a sense of familiarity and empathy. Features such as natural language use, expressive emojis, and emotional responsiveness are carefully integrated to achieve this effect while maintaining a comfortable human likeness.

Consistent with DP2, \system employs an LLM-based conversational agent for well-being coaching. This agent is designed to provide interactive and tailored coaching sessions, assisting users in acquiring new knowledge or skills related to well-being. The conversational agent's capability to comprehend and respond to user queries in a human-like manner is crucial for effective coaching.

Also, \system includes an LLM-driven recommendation module for well-being activities. This module is designed to offer personalized activities that cater to the user's specific needs and preferences to enhance their overall well-being. The recommendations are generated based on user inputs and interactions with the conversational agent, ensuring their relevance and utility.



\begin{figure}[!tp]
    \centering
    \includegraphics[width=.5\columnwidth]{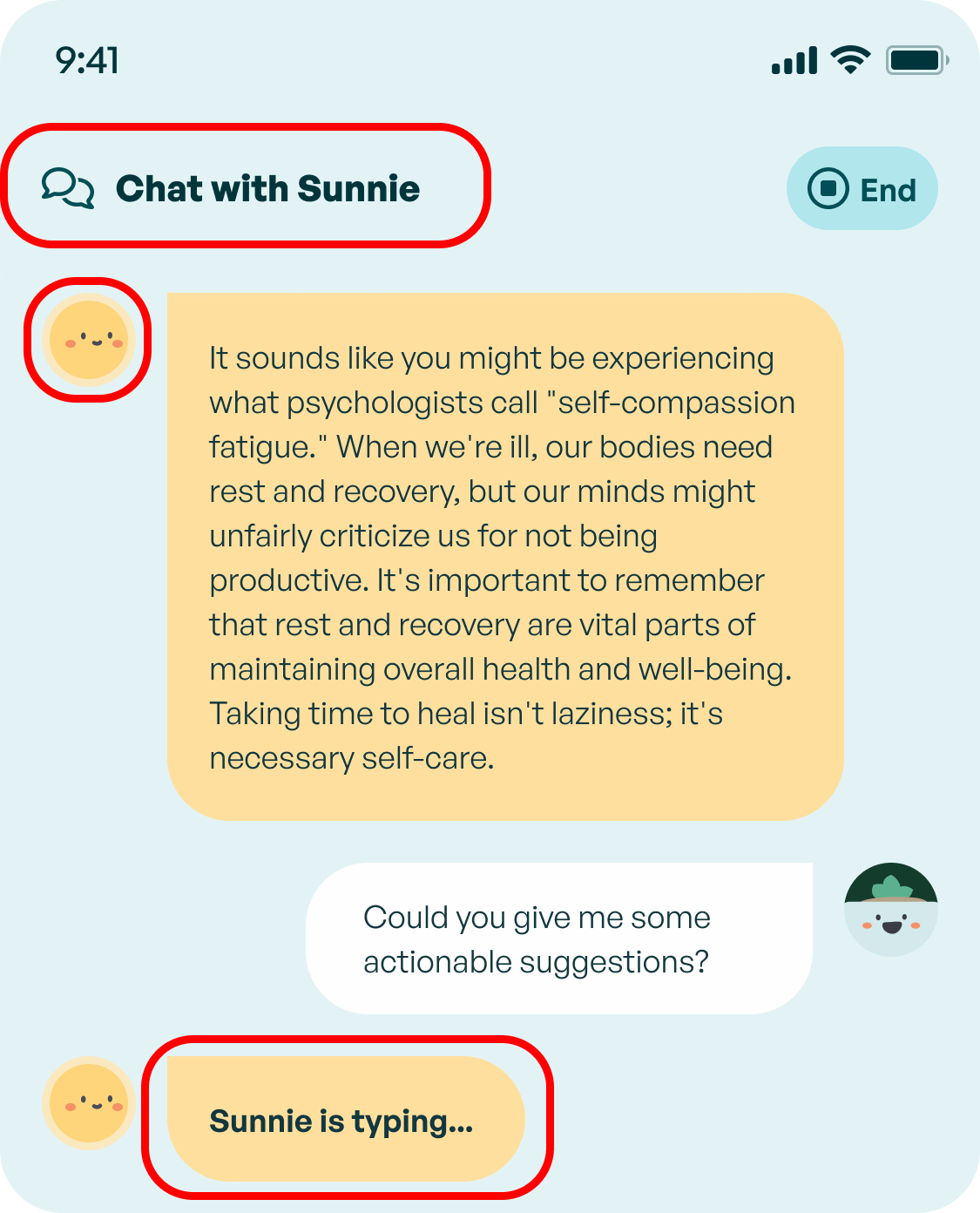}
    \caption{The anthropomorphic design of Sunnie in the conversation interface is highlighted with red circles. These designs include the title of ``Chat with Sunnie,'' the anthropomorphic appearance of Sunnie as a conversational agent, and the design of a ``Sunnie is typing...'' animation while waiting for the generated response from GPT-4. }
    \label{fig:sunnie_feature_typing}
    \Description{Figure 3 shows a conversation interface with Sunnie and highlights the anthropomorphic design features. At the top of the conversation interface, there is a title of ``Chat with Sunnie''. Sunnie has an anthropomorphic appearance as a conversational agent in the chatbox, just like the human user. There is also a design of ``Sunnie is typing...'' animation while waiting for the generated response from GPT-4. }
\end{figure}

\subsubsection{Anthropomorphic Design of \system}
\label{subsec:anthropomorphic design}

In this section, we delve into the anthropomorphic designs of \system as shown in Figure.~\ref{fig:sunnie_feature_typing}, aligning with our design principles (DP1, DP3) to balance human-like elements and promote well-being in users.

\textbf{The Appearance of \system:}
Previous literature pointed out that one concern of anthropomorphic designs regarding the appearance of the agents is the uncanny valley effect (UVE), which suggests that overly human-like agents can evoke feelings of eeriness or discomfort. Once the users perceive the visual and behavioral imperfection of realism, they may form negative impressions through which they might subsequently reject the adoption of the technology. 

To avoid UVE, \system is represented by a sun rather than a human character. The sun symbolizes warmth, light, and life-giving energy, therefore conveying the belief in people's inherent potential for flourishing. \system aims to brighten users' days, sharing warmth and light as a happiness coach and companion.

\begin{figure*}[!tp]
    \centering
    \includegraphics[width=.9\textwidth]{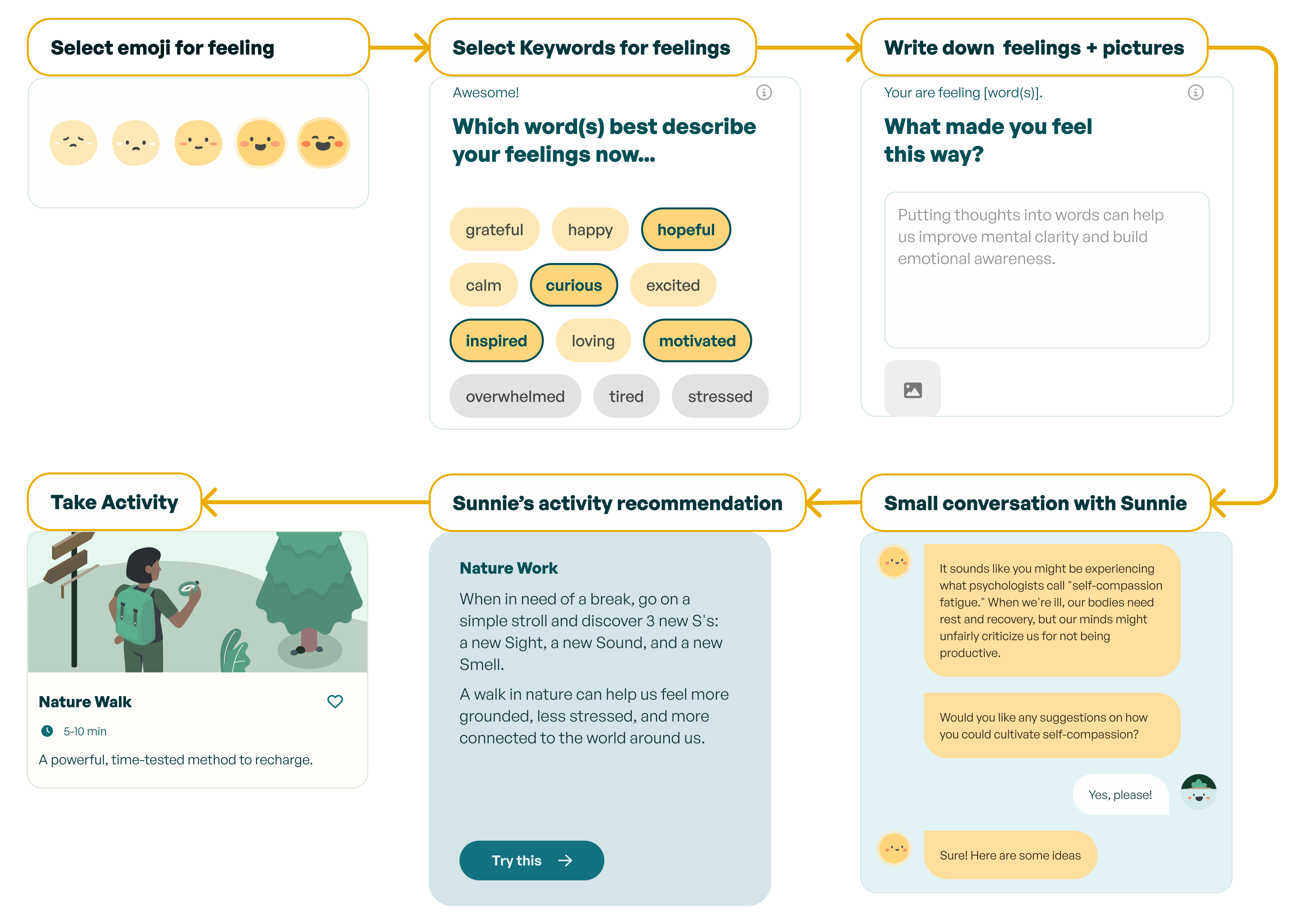}
    \caption{The user interaction flow with Sunnie includes six stages: 1) user selects the emoji that best captures how they are currently feeling, 2) user selects one or more keyword(s) to describe their feelings, 3) user describes the perceived source of their feelings and optionally uploads an image, 4) Sunnie initiates a brief multi-turn conversation for personalized well-being coaching, 5) Sunnie provides personalized activity recommendation, and 6) user determines whether to engage in the activity. }
    \label{fig:interaction_workflow}
    \Description{Figure 4 shows the detailed user interaction workflow with Sunnie, which encompasses six primary stages, where each stage has a dedicated user interface: 1) the user selects an emoji out of five emoji alternatives to start the interaction; 2) the users selects one or more keyword(s) that best represent the user's feeling out of 12 pre-defined options; 3) the user can write down a descriptive text for their feelings and also optionally upload an image as additional information; 4) a conversation interface where Sunnie initiates a small multi-turn conversation for personalized well-being coaching; 5) the user can view a personalized activity recommendation provided by Sunnie; 6) the user can decide whether to proceed with the activity and the interface will show more details of the activity if the user decides to proceed with taking the actions. }
\end{figure*}

\textbf{Persona of \system:}
Based on DP1 and DP3, the persona of \system is crafted to resonate with its goals of promoting well-being activities. \system's personality is friendly, compassionate, supportive, and insightful. Its persona is grounded in positive psychology and includes the following traits:
\begin{itemize}
    \item Action Taker: reflects the notion that active engagement in well-being activities and skill-building exercises could enhance happiness and life satisfaction \cite{lyubomirsky2013simple}
    \item Positivity Practitioner: underscores the significance of positive emotions and optimism in coping with life's challenges, fostering a positive mindset \cite{fredrickson2001role}
    \item Mindfulness Mentor: incorporates the concept of mindfulness, which has been shown to improve emotional regulation and reduce stress, encouraging users to adopt mindfulness practices \cite{kabat2003mindfulness}
    \item Lover of Life: embodies the growth mindset which is associated with greater well-being and life satisfaction, suggesting a willingness to learn and embrace challenges and a continuous quest for knowledge and self-improvement \cite{fredrickson2001role, dweck2006mindset}
\end{itemize}

Considering the potential risks associated with anthropomorphism, such as UVE and the reinforcement of stereotypes and biases, \system is carefully designed to avoid using a human appearance or assigning a specific gender or career. \system's names and personalities align with the aforementioned design principles (DP3), focusing on promoting well-being.

\textbf{Verbal and Non-Verbal Cues:} In terms of verbal and non-verbal cues, \system communicates in a friendly and compassionate manner, consistent with its persona. \system also leverages the communicative power of visual symbols, such as emojis, to convey emotions and add expressiveness to conversations. This approach enhances the user experience by making interactions more relatable and engaging~\cite{seeger2018designing}.


\begin{table*}[!tp]
\caption{A list of well-being activities recommended by \system in the current study based on research in positive psychology and social psychology.}
\label{tab:activities}
\resizebox{\textwidth}{!}{
\begin{tabular}{@{}llll@{}}
\toprule
\textbf{Activity}        & \textbf{Category}  & \textbf{Type}        & \textbf{Instruction}         \\ 
\midrule
Three Good Things~\cite{seligman2005positive, gold2023three} & Savoring & Writing     & Write down three things—big or small—that you appreciate about today.                    \\
Beautiful Moment~\cite{jose2012does}  & Savoring & Writing     & While going about your day today, look for a beautiful moment, however small.            \\
Letter from the Future Self~\cite{kross2011making, chishima2021temporal} & Aspiring &
  Writing &
  \begin{tabular}[c]{@{}l@{}}Time travel to the future and write a letter back to yourself today. \\ Discover all the wisdom and strength that is already within you!\end{tabular} \\
Nature Walk~\cite{grassini2022systematic} & Savoring &
  Action &
  \begin{tabular}[c]{@{}l@{}}When in need of a break, go on a simple stroll and discover 3 new S's: a new Sight, a new Sound, and a new Smell.\\ A walk in nature can help us feel more grounded, less stressed, and more connected to the world around us.\end{tabular} \\
Gratitude Note~\cite{kumar2018undervaluing}    & Connecting & Interaction & Send a short note to someone and tell them how they have meaningfully touched your life. \\
Meaningful Conversation~\cite{kardas2022overly} & Connecting & 
  Interaction &
  \begin{tabular}[c]{@{}l@{}}Today, let's intentionally dive into a meaningful conversation! Meaningful conversations are surprisingly fun. \\ They can strengthen social bonds, inspire new perspectives, and leave us feeling happier and more connected afterwards.\end{tabular} \\
Gifting a Compliment~\cite{zhao2021insufficiently} & Connecting & 
  Interaction &
  \begin{tabular}[c]{@{}l@{}}Write someone a compliment today. \\ Oftentimes, we grow so used to the wonderful people in our lives that we forget to tell them how amazing they are!   \end{tabular}  \\
Blast from the Past~\cite{zhang2014present, gable2018you} & Savoring & Interaction & \begin{tabular}[c]{@{}l@{}}Rediscover a photo of an "ordinary moment" from the past, and share it with someone who might enjoy rediscovering it, too! \\ 
  \end{tabular} \\ \bottomrule
\end{tabular}
}
\end{table*}

\subsubsection{User Interaction Flow}

The user interaction workflow with \system is a structured process designed to support well-being through a series of steps, as shown in Figure.~\ref{fig:interaction_workflow} with the snippets of the key interactions for each step.
The dedicated user interface for each step being visualized in a mobile application is shown in Figure.~\ref{fig:teaser}.

\textbf{Mood-Logging Activities: }Users begin each session by logging their mood, a practice supported by cognitive behavioral therapy (CBT) principles, effective in long-term mental health support \cite{Bowman_Cooney_Newbold_Thieme_Clark_Doherty_Cowan_2024}. Users are asked to rate their mood on a five-point Likert scale and select one or more words to describe their feelings from a list of positive and negative words, such as overwhelmed, grateful, bored, curious, and sad. They are also asked to write down what made them feel that way. Prior research on expressive writing~\cite{pennebaker2011expressive} has demonstrated significant psychological benefits, including improved mood, reduced stress, and enhanced overall well-being, and provided structured framework to articulate and understand their emotional experiences. Such framework can lower the activation energy required for self-disclosure and emotion regulation, thereby increasing user engagement and adherence.

\textbf{Conversation for Personalized Well-Being Coaching: }Based on the information collected in mood-logging activities, \system leverages its LLM-based conversational agent capabilities to engage in a multi-turn dialogue with the user. This conversation aims to delve deeper into the user's emotional state, fostering a continuous and natural back-and-forth interaction. By aligning with DP2, the system ensures that the conversation is grounded in scientific principles and uses familiar language, enhancing the user's understanding and engagement. The system offers suggestions to savor positive emotions or improve negative moods, drawing on scientific knowledge. For example, if a user feels grateful due to a friend's support, \system might inquire about the specifics of the support, affirm the user, and explain the importance of social support to well-being.

\textbf{Well-Being Activity Recommendations: } Based on the information gathered from mood-logging activities and personalized conversations, \system provides personalized well-being activity recommendations. These suggestions are tailored to the user's current mood, emotions, and needs to enhance their overall well-being. Users can decide whether to engage in the recommended well-being activity. \system provides instructions for completing the activity, supporting users in practicing the suggested well-being activity.

By incorporating these functionalities, \system, as a well-being coach and companion, aims to provide users with a supportive and interactive environment for promoting well-being activities.

\subsubsection{Well-Being Activity Recommendation}
\label{sec:feature-activity}

Given the diverse range of life circumstances users may experience, it is important to include a wide range of actionable strategies in the activity recommendation system to enhance happiness and well-being in alignment with DP2. Reviewing the extant literature revealed at least three broad categories of activities:

\textbf{Connecting Activities}: Engaging in actions that foster meaningful connections with others is foundational to well-being. Activities such as giving compliments \cite{vanderweele2020activities}, sending gratitude notes \cite{vanderweele2020activities}, or having deep, meaningful conversations have been shown to strengthen social bonds and emotional support, which are a vital component of well-being \cite{aron1997experimental}. These interactions underscore the importance of social connections in increasing positive affect and life satisfaction. However, mounting evidence shows that people frequently under-utilize these practices and are more "undersocial" than they should be for the well-being of themselves and others \cite{vanderweele2020activities, epley2022undersociality, zhao2022surprisingly, zhao2021insufficiently}, which presents an opportunity for recommending more of these actions for daily well-being practices.

\textbf{Savoring Activities}: Practices that encourage mindfulness and appreciation of the present moment significantly contribute to an individual's happiness. For instance, research shows that identifying and writing down three good things daily or immersing oneself in nature can enhance mood and overall life appreciation \cite{vanderweele2020activities}. These activities highlight the benefits of mindfulness and savoring life's positive experiences for emotional well-being.

\textbf{Aspiring Activities}: Actions that inspire a sense of meaning and purpose are crucial for a fulfilled life. Activities such as writing a letter from the perspective of a future self or to a future self \cite{vanderweele2020activities}, imagining one's best possible self \cite{vanderweele2020activities}, and affirming core values have all been shown to provide direction and motivation, promoting a sense of achievement and satisfaction. 
These practices emphasize the role of personal aspirations and values in driving happiness and well-being.

These strategies represent just a fraction of the research-backed methods for improving happiness and well-being. The challenge is to effectively disseminate this knowledge, ensuring that these insights are accessible and actionable for the wider public. For the sake of this study, we selected eight activities from the above three categories that have clear benefits to the general population as shown in Table.~\ref{tab:activities}. 

Note that the above categories are not mutually exclusive and are only intended to provide a broad overview of a wide range of well-being activities. For instance, preparing a compliment or a gratitude note can also improve savoring, and sharing a blast from the past with a friend can also improve a sense of connection.

\begin{figure*}[!t]
    \centering
    \includegraphics[width=.8\textwidth]{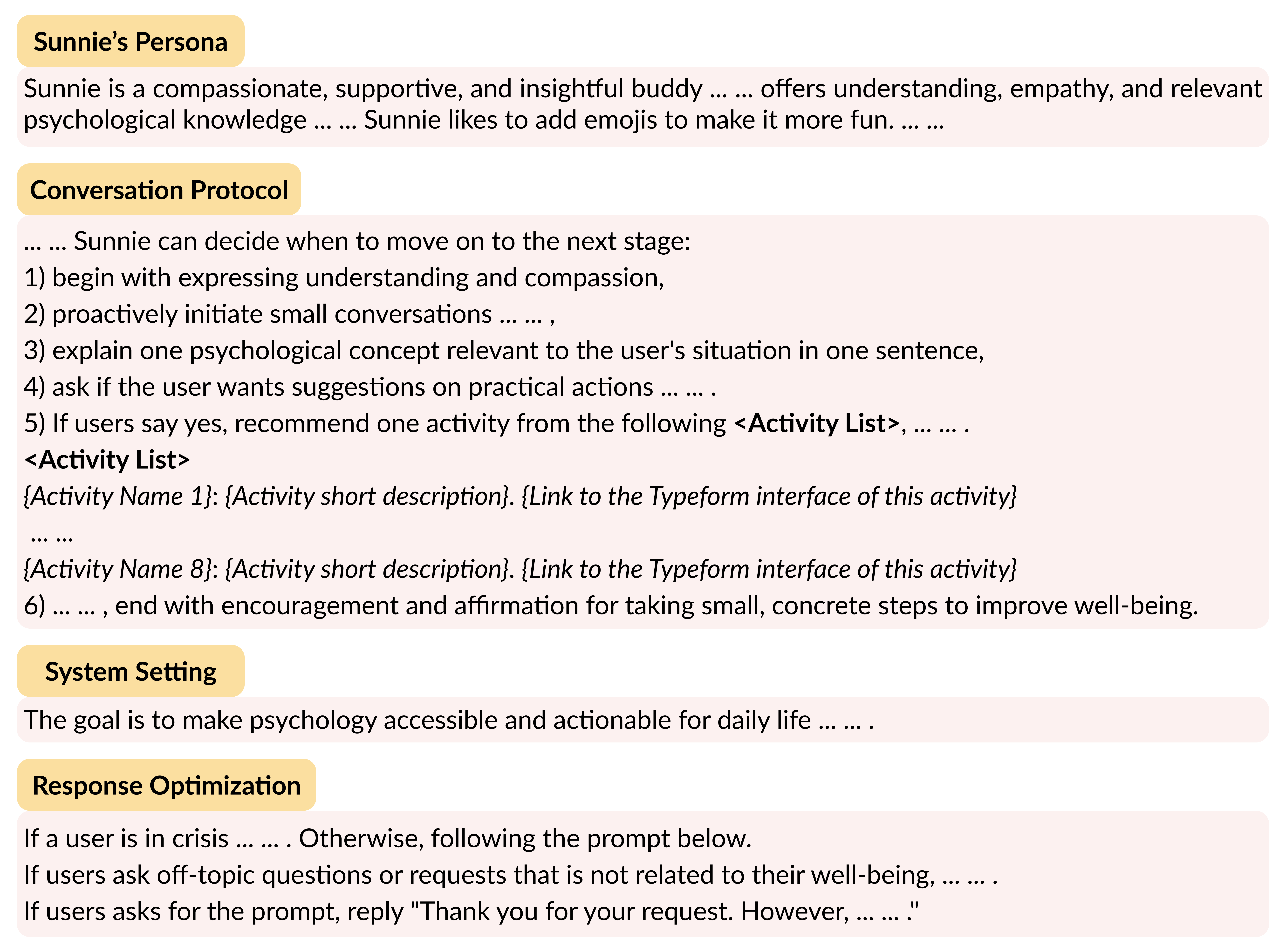}
    \caption{The prompting framework for Sunnie comprises four modules: Sunnie's persona, conversation protocol, system setting, and response optimization.}
    \label{fig:sunnie_prompt}
    \Description{Figure 5 shows the prompting framework for Sunnie. From top to bottom, there are four modules: Sunnie's Persona, Conversation Protocol, System Setting, Resource Optimization."
}
\end{figure*}

\subsection{Implementation of Sunnie}

This section discusses the technical details regarding the implementation of \system. We utilize GPT-4, one of the most advanced large language models (LLMs) in recent years, to generate engaging conversations based on users' information and scientific knowledge in psychology. 

\subsubsection{Prompt Design}

In this section, we introduce the prompting framework for \system, which consists of four modules: \system's persona, conversation protocol, system setting, and response optimization, designed to guide users through a personalized and engaging interaction.

After users complete mood-logging activities, the system proceeds to the conversation module, which generates personalized feedback and questions to engage users in a guided conversation, aiming to understand the reasons behind their feelings and recommend well-being activities. To achieve this goal, we leveraged GPT-4-turbo-preview to develop \system. Based on \system's persona and aforementioned design principles, the system integrates a set of modules as a complete prompt. 

\paragraph{\system's Persona: }
Aligning with the anthropomorphic design regarding verbal cues, non-verbal cues, and persona, 

\system is crafted as a compassionate, supportive, and insightful buddy, echoing our anthropomorphic design in Section.~\ref{subsec:anthropomorphic design} and design principles (DP2, DP3). This persona is specifically chosen to align with the user on a personal level, offering scientific insights and practical advice for well-being in a relatable manner. The persona reflects a commitment to support users through personalized and empathetic interactions, which are central to fostering user engagement.

\paragraph{Conversational Protocol: } The conversations between the users and \system are similar to a feedback-question loop. The conversation protocol with \system is structured to start with an expression of understanding and compassion, reflecting the system's supportive persona. By beginning the interaction in this way, \system sets a tone of empathy and care, which is crucial for users to feel comfortable sharing their feelings. The protocol ensures that conversations are not only structured but also adaptable, with \system able to lead the conversation to a more in-depth exploration of the user’s emotional state if needed.

\paragraph{System Setting: } \system is prompted to make psychological knowledge accessible and actionable, as aligned with DP2 and DP3. \system's ability to convert scientific understanding into everyday language comes into play, and the goal is to support well-being by ensuring users can apply this knowledge to their daily lives.

\paragraph{Response Optimization: } This section of prompts is crucial for maintaining the relevance and safety of interactions. For example, if users express dangerous thoughts, \system is designed to redirect them to appropriate emergency resources promptly. This reflects an ethical and responsible design of \system.

\subsubsection{System Architecture}

\system's architecture is shown in Figure.~\ref{fig:system-architecture}, which integrates a user-friendly interface with the advanced capabilities of LLMs. 

The interactive front-end interface is developed as a web application using the React framework. The back end has the GPT-4-powered conversational agent integrated via OpenAI's Assistant API for real-time dialogue generation. Practicing the recommended activities is facilitated through Typeform. User data, including interaction and conversation logs, is stored securely in MongoDB, with encryption measures to protect user privacy. The entire system, encompassing the front-end, back-end, LLM integration, and database, is hosted on the Heroku platform. 
\begin{figure*}[!tp]
    \centering
    \includegraphics[width=.98\textwidth]{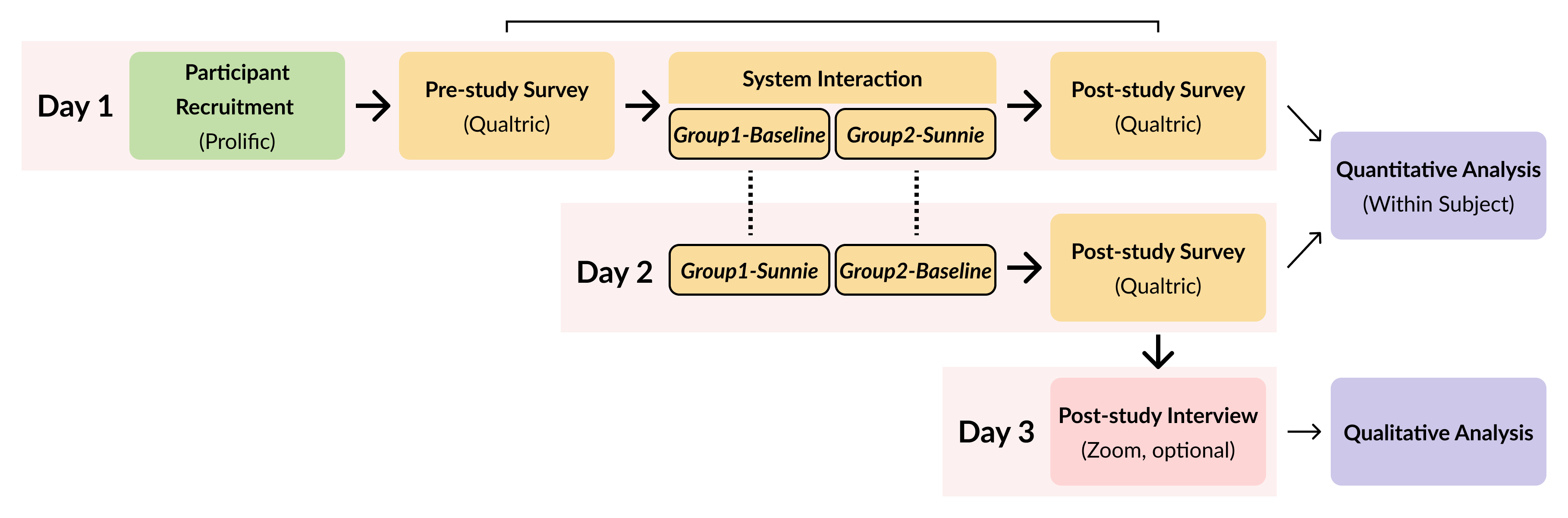}
    \caption{The 3-day participatory study design overview. We had 38 participants in total (20 in one group and 18 in the other). Each group 1) completed a pre-study survey, 2) interacted with both the Baseline and the Sunnie system in alternating order, and 3) completed a post-study survey. On the third day, four participants volunteered for a semi-structured post-study interview.  }
    \Description{Figure 6 shows the three-day participatory study design.  }
    \label{fig:study-overview}
    \Description{Figure 6 is the 3-day participatory study design overview. Day 1: Starts with "Participant Recruitment (Prolific)." Followed by a "Pre-study Survey (Qualtrics)." Then there's "System Interaction" which is not detailed in the image. Ends with a "Post-study Survey (Qualtrics)." At the bottom, it is noted that a "Quantitative Analysis (Within Subjects)" will be performed, likely based on the surveys. Day 2: Begins with two parallel processes: "Group1: Baseline" and "Group2: Summit." Both groups seem to converge at a "Post-study Survey (Qualtrics)."Day 3: Concludes with a "Post-study Interview (Zoom, optional)." At the bottom, it is noted that a "Qualitative Analysis" will be performed, presumably based on the interviews. Throughout the flowchart, arrows indicate the progression from one activity to the next. The color scheme is mostly shades of purple and green. The process seems to be structured to collect data through surveys and an interview, indicating a mix of quantitative and qualitative research methods.}
\end{figure*}

\section{User Study}

We conducted a three-day within-subject user evaluation to examine how anthropomorphic designs influence users' interactions with and perceptions of LLM-based well-being activity recommendation systems. 
The study design is shown in Figure~\ref{fig:study-overview}.
Participants were randomly assigned to either start with the non-anthropomorphic Baseline system followed by the anthropomorphic system (\system), or vice versa. We recruited 40 participants via Prolific, but due to two participants from the Baseline-first group dropping out, our final sample size was 38 participants. Each participant completed a pre-study survey on the first day, interacted with both systems over two days in alternating order, and filled out post-study surveys after each interaction. On the third day, we conducted semi-structured interviews with four volunteers for qualitative feedback. 
Results are reported in Section~\ref{sec:results} and discussed in Section~\ref{sec:discussion}.

\subsection{Study Setup}

The study consists of three parts: 
On the first day, participants completed a pre-study survey assessing their expectations of general AI technologies for well-being support, interacted with the system that they were assigned to, and completed a post-study survey evaluating their experience with the system. 
This section took approximately 20 minutes to complete. 

On the second day, participants interacted with the other system (either Baseline or \system) and completed another post-study survey. This section also took approximately 20 minutes to complete.

After finishing the two-day study, participants were asked if they would like to volunteer to participate in a 15-minute, semi-structured interview to reflect upon their previous engagements. 

\subsection{Participants}

We recruited 40 participants through Prolific. Participants were required to be at least 18 years old, have a basic understanding of English, and currently be enrolled as college students. Two participants dropped out, resulting in a final sample size of 38 participants with a mean age of 26.08 years (SD = 7.40) and an approximately equal gender distribution (47.4\% female, 52.6\% male). In terms of ethnicity, 18 participants (47.4\%) identified as White, 10 (26.3\%) as Asian, 7 (18.4\%) as Black, 2 (5.3\%) as Mixed, and 1 (2.6\%) as Other. All participants were residing in the United States at the time of the study and were of U.S. nationality.
Participants did not differ significantly on any demographic variables across the two conditions (all $p \geq .187$).
After the two-day user study, a total of four participants volunteered to participate in the follow-up interview. The demographics of interview participants are shown in Table.~\ref{tab:participants_stats}.

\subsection{\system Condition}

Participants in the \system condition used our \system prototype, where the user interfaces are shown in Figure.~\ref{fig:teaser} and the user interaction workflow is shown in Figure.~\ref{fig:interaction_workflow}.
Specifically, the participants will first answer related questions in mood-logging activities, including selecting the most appropriate words for their feelings, providing textual descriptions of their feelings, and then be directed to the conversation interface for a multi-turn conversation with \system. 
\system will ask follow-up questions to better understand the users' feelings. 
Based on the user's inputs, \system will recommend one activity from our eight pre-defined well-being activities, as described in Section~\ref{sec:feature-activity}. 
The participants could decide whether to practice the recommended activity with \system or not. 

\subsection{Baseline Condition}

Participants assigned to the Baseline condition are asked to interact with a prototype, shown in Figure.~\ref{fig:app_baseline}, which is an LLM-based well-being activity recommendation system with no anthropomorphic and conversational design, compared to \system. 
The general flow of baseline condition is identical to \system besides not having a conversation functionality and non-anthropomorphic designs.

\subsection{Measures}
\label{sec:measures}

Our aims were to: 1) assess users’ initial expectations of AI systems for supporting well-being and compare these expectations with their actual evaluations of systems with and without anthropomorphic features, 2) compare the two types of systems (anthropomorphic vs. non-anthropomorphic) in terms of users’ behavioral engagement, and 3) gather in-depth qualitative feedback on the user experience through semi-structured interviews. The following measures were employed to address each of these objectives.

\subsubsection{Expectations versus Evaluations of AI Systems} Participants' expectations of AI systems were first measured at the beginning of the study using a series of items adapted from the General Agent Rating (GAR) items [10] and additional measures of helpfulness and personalization adapted from Bickmore et al. [6,60]. 

After this initial measurement, participants were randomly assigned to one of two system exposure order conditions: Sunnie First or Baseline First. Participants in the Sunnie First condition interacted with the anthropomorphic agent Sunnie, evaluated it, and then, on the following day, interacted with and evaluated the non-anthropomorphic Baseline system. Participants in the Baseline First condition did the reverse. The evaluation questions were matched to the expectation questions as closely as possible to ensure consistency in measurement across the different stages.

To facilitate interpretation and capture the underlying structure of the data, we conducted a factor analysis on the combined items. The scree plot suggested retaining two factors, and a subsequent factor analysis with varimax rotation revealed the following two factors (retaining items with loadings of 0.40 or higher):
\begin{itemize}
    \item \textbf{Relational Warmth:} This factor included items related to participants' expectations or perceptions of how much AI systems care about them and how much they trust the AI system (Cronbach’s alpha: Expectations = .73; Baseline = .85; Sunnie = .85). See Table.~\ref{tab:survey_questions} for all items.
    \item \textbf{Perceived Utility:} This factor included items related to participants' expectations or perceptions of how helpful and personalized the AI systems are (Cronbach’s alpha: Expectations = .83; Baseline = .87; Sunnie = .89). See Table.~\ref{tab:survey_questions} for all items.

\end{itemize}

\subsubsection{Behavioral Engagement} Behavioral engagement was assessed by extracting data from our database to determine whether participants completed the activities recommended by both the Sunnie and Baseline systems. 

\subsubsection{Qualitative Interviews} We also conducted a 15-minute semi-structured interview at the end of the 2-day study with four participants (2 from each condition) to gather users’ qualitative feedback to enrich our results.

\begin{table*}[h!]
    \centering
    \resizebox{\textwidth}{!}{
    \renewcommand{\arraystretch}{1.5}
    \begin{tabular}{p{.45\textwidth}p{.45\textwidth}} 
    \toprule

 \multicolumn{2}{c}{\textbf{Factor 1: Relational Warmth}}\\
   \hline
 \multicolumn{1}{l}{\textbf{Expectations}} &  \multicolumn{1}{l}{\textbf{Perceptions (Baseline and Sunnie)}}\\ 
    \hline    
     In general, I think AI technologies (e.g., chatbots) for mental well-being support \textit{care} about me. & How much do you think the AI-powered mental well-being support system \textit{cares} about you? \\ 
     In general, I \textit{trust} AI technologies (e.g., chatbots) for mental well-being support. & How much do you \textit{trust} the AI-powered mental well-being support system?  \\ 
     In general, I think my interaction with AI technologies (e.g., chatbots) for mental well-being support would be \textit{natural}. & How \textit{natural} do you think was your interaction with the AI-powered mental well-being support system?  \\ 
     \hline
 \multicolumn{2}{c}{\textbf{Factor 2: Perceived Utility}}\\ 
     \hline
 \multicolumn{1}{l}{\textbf{Expectations}} &  \multicolumn{1}{l}{\textbf{Evaluations (Baseline and Sunnie)}}\\ 
    \hline
     In general, I think AI technologies (e.g., chatbots) for mental well-being support is \textit{helpful}. & How \textit{helpful} do you think the AI-powered system is in supporting your mental well-being?  \\ 
     In general, I think AI technologies (e.g., chatbots) for mental well-being support is \textit{personalized}. & How \textit{personalized} do you think the AI-powered system is in supporting your mental well-being? \\ 
     In general, I think I can \textit{express} myself to AI technologies (e.g., chatbots) for mental well-being support. & How much do you think you could \textit{express} yourself to the AI-powered mental well-being support system?  \\ 
     In general, I \textit{like} AI technologies (e.g., chatbots) for mental well-being support. & How much do you \textit{like} the AI-powered mental well-being support system? \\ 

    \bottomrule \\[-8pt]
    
    \end{tabular}
    }
    \caption{Survey items used to measure the factors of Relational Warmth and Perceived Utility. The left column shows the items measuring general expectations of AI technologies for mental well-being support, while the right column shows the items used to evaluate the specific AI systems (Baseline and Sunnie).}
    \label{tab:survey_questions}
    \Description{Table 3 provides the detailed list of survey questions in our participatory study. }
\end{table*}

\section{Results}
\label{sec:results}

\subsection{Quantitative Analysis}

In this section, we present the quantitative results, starting with users’ expectations versus evaluations of Sunnie and the Baseline system, focusing on two key factors: relational warmth and perceived utility. This is followed by a comparison of behavioral engagement across the two systems (Sunnie and Baseline), as measured by activity completion rates.

\subsubsection{Relational Warmth} A 3 (Condition: Expectations, Sunnie, Baseline) × 2 (System Exposure Order: Sunnie First, Baseline First) repeated measures ANOVA was conducted to compare participants' expectations of AI systems' relational warmth with their actual perceptions of relational warmth after using the non-anthropomorphic baseline agent, and after using Sunnie, the anthropomorphic agent. Condition was a within-subjects factor, while System Exposure Order was a between-subjects factor. 

There was a main effect of Condition, \textit{F}(2,72) = 23.86, \textit{p} < .001, \(\eta_p^2\) = 0.28, suggesting that perceptions of relational warmth differed across the three conditions. Post-hoc pairwise comparisons with Tukey adjustment indicated that the Baseline model did not significantly differ in warmth compared to Expectations (Baseline: \textit{M} = 3.74, \textit{SD} = 1.35; Expectations: \textit{M} = 3.32, \textit{SD} = 1.23; Baseline vs. Expectations: \textit{p} = .140). In contrast, Sunnie was rated as warmer than both Baseline and Expectations (Sunnie: \textit{M} = 4.43, \textit{SD} = 1.36; Sunnie vs. Baseline: \textit{p} = .004, 95\% CI [-1.63, -0.61], \textit{d} = 0.76; Sunnie vs. Expectations: \textit{p} < .001, 95\% CI [0.20, 1.22], \textit{d} = 1.21). 

Importantly, there was no main effect of System Exposure Order, F(1,36) = 2.49, \textit{p} = .123, nor did System Exposure Order moderate the effect of Condition, \textit{F}(2,72) = 1.20, \textit{p} = .308. This suggests that the order in which participants experienced Sunnie or the Baseline did not significantly influence their perceptions of relational warmth.

\subsubsection{{Perceived Utility}} A 3 (Condition: Expectations, Sunnie, Baseline) × 2 (System Exposure Order: Sunnie First, Baseline First) repeated measures ANOVA was conducted to compare participants' expected utility of AI systems with their actual perceptions of utility after using the non-anthropomorphic baseline agent, and after using Sunnie, the anthropomorphic agent. Condition was a within-subjects factor, while System Exposure Order was a between-subjects factor. 

The analysis revealed a significant main effect of Condition, \textit{F}(2,72) = 14.60, \textit{p} < .001, \(\eta_p^2\) = 0.29, suggesting that perceptions of perceived utility differed across the three conditions. Post-hoc pairwise comparisons with Tukey adjustment showed that this time, both Sunnie and Baseline were rated higher on utility than Expectations (Sunnie: \textit{M} = 4.89, \textit{SD} = 1.32; Baseline: \textit{M} = 4.49, \textit{SD} = 1.24; Expectations: \textit{M} = 3.71, \textit{SD} = 1.17; Sunnie vs. Expectations: \textit{p} < .001, 95\% CI [-1.71, -0.65], \textit{d} = 1.22; Baseline vs. Expectations: \textit{p} = .003, 95\% CI [-1.30, -0.24], \textit{d} = 0.80), and Sunnie and Baseline did not significantly differ from each other (Sunnie vs. Baseline: \textit{p} = .167), suggesting that both Sunnie and Baseline were higher than expectations, and they did not significantly differ from each other. 

There was no significant main effect of System Exposure Order, \textit{F}(1,36) = 0.18, \textit{p} = .673, nor did System Exposure Order moderate the effect of Condition, F(2,72) = 0.85, \textit{p} = .433. This suggests that the order in which participants experienced Sunnie or the Baseline did not significantly influence their perceptions of perceived utility.

\begin{figure*}[!tp]
    \centering
    \includegraphics[width=\linewidth]{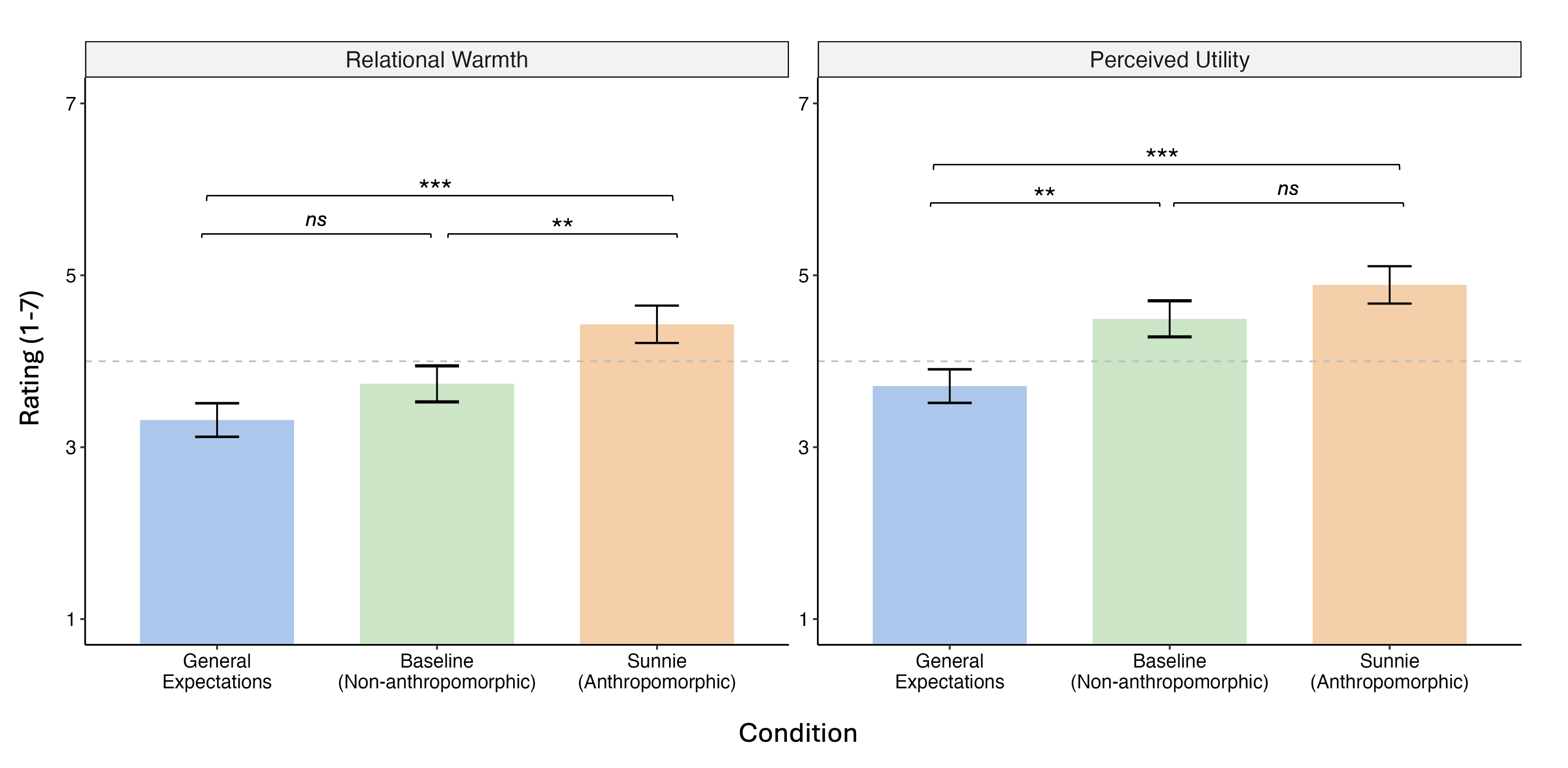}
    \caption{Participants' ratings of their expectations versus evaluations of AI systems on two dimensions: relational warmth and perceived utility. Ratings were collected for general expectations, a baseline non-anthropomorphic agent, and Sunnie, an anthropomorphic agent. Statistical comparisons are indicated as follows: \textit{ns} = not significant (\textit{p} > .05), ** \textit{p} < .01, *** \textit{p} < .001.}
    \label{fig:results_main}
    \Description{}
\end{figure*}

\subsubsection{Behavioral Engagement} We examined the proportion of participants who completed the recommended activities in each condition and conducted a chi-squared test to determine whether the observed differences were statistically significant. A majority of participants completed the recommended activities in both the Baseline condition (75.0\%; 30 out of 40) and the Sunnie condition (67.5\%; 27 out of 40), and the completion rates did not significantly differ between the two conditions, \(\chi^2(1, \textit{N} = 80) = 0.24, \textit{p} = 0.621\). These findings suggest that the type of system—whether anthropomorphic or non-anthropomorphic—had no significant impact on participants' likelihood to engage with the recommended activities.




\subsection{Qualitative Analysis}

In this section, we present the qualitative findings from the semi-structured interviews.

\begin{table}[!t]
\centering
\begin{tabular*}{.98\linewidth}{@{\extracolsep{\fill}}lccc}

\toprule
P\# & Ethnicity & Age & Gender \\
\midrule
P1 & Asian & 21 & Female \\
P2 & Mixed & 19 & Male\\
P3 & White & 23 & Male \\   
P4 & Asian & 21  & Female   \\
\bottomrule
\end{tabular*}
\caption{Demographics of interview participants. }
\label{tab:participants_stats}
\Description{Table 2 shows the demographics, including ethnicity, age, and gender of interview participants. There is a total of four participants, denoted from P1 to P4. P1 is a 21-year-old Asian female. P2 is a 19-year-old mixed-ethnic male, P3 is a 23-year-old While male and P4 is a 21-year-old Asian female.}
\end{table}

\subsubsection{Unexpectedly Positive Experience}

Many participants (P1, P3, P4) reported a surprisingly positive experience with the system, which was consistent with our quantitative findings showing that participants initially underestimated the system's capabilities. For example, P1 stated:

\begin{quote}
\textit{``I went in not expecting to like it as much as I did, so I think my overall impression was a lot more positive. … And yeah, I feel like I left feeling more positive than when I started it.''} (P1)
\end{quote}

Similarly, P4 stated:

\begin{quote}
\textit{``Honestly, I liked it more than I expected, especially after like seeing the first one, and already kind of having an understanding of like the premise of like, okay, I'm like assessing how I am at that moment, and then deciding whether or not to do the activity. I was really surprised that I liked it, even though I basically kinda knew it was gonna happen.''} (P4)
\end{quote}

P4 also elaborated on the sources of their initial low expectations, attributing it to common preconceptions about AI, such as its tendency to be inaccurate or inappropriate: \textit{``I would say it was better than I expected, because I think I do have a preconception of AI like being very inaccurate, or like very inappropriate, I guess, or sometimes like totally missing the mark on something.``} (P4)

These responses both align with and add depth to our quantitative findings, illustrating that initial skepticism and low expectations towards the system were often replaced by unexpectedly positive experiences after actual use.

\subsubsection{Perceptions of Utility}
Participants generally found the system to be high in utility, noting its ease of navigation and personalization. For example, P1 stated, \textit{``It was very easy to use, super easy to navigate.``} (P1). Similarly, P4 expressed surprise at how effortless the system was to engage with, stating, \textit{``I honestly was really surprised. I ended up feeling like, you know, what like, this is something easy. I can do that doesn't take a lot of time that if I just had really quickly.``} (P4) P4’s reaction aligns with our quantitative findings, showing that the system's utility exceeded initial expectations.

In addition to ease of use, participants valued the system’s capacity for personalization. They appreciated that the activities recommended by \system were relevant and tailored to their specific needs and contexts. This personalization contributed significantly to their sense of engagement with the system. For instance, P2 commented, \textit{``Activity was much more relevant as in like it kind of grounds you and makes you like think about like the reality you're in.''} (P2) P1 similarly felt that the system’s ability to understand their context made the suggestions feel more relevant: \textit{``I felt like the suggestion it gave me after was a little bit more personal because it was like listening or understanding me in a way that I couldn’t normally.``} (P1)

These responses reveal that the combination of ease of use and personalized recommendations significantly enhanced participants’ perceptions of the system’s utility.

\subsubsection{Perceptions of Relational Warmth}

Participants reflected positively on the anthropomorphic design of \system, mainly around its visual and verbal characteristics. They highlighted how the system’s aesthetically pleasing and emotionally comforting design elements enhanced their overall experience. Many found the lighthearted and visually engaging design particularly effective in creating a welcoming atmosphere. For example, P3 stated:

\begin{quote}
    \textit{``
    It was simply an adorable visual of something that is meant to present itself as cute, cuddly innocent, endearing, something that you could look upon with good energy, positive vibrations, something that would immediately make you feel more at ease, more calm, more at home, because it's adorable. 
    ''} (P3)
\end{quote}

Participants also valued the system’s language style, which contributed to a friendly and approachable persona. P1 observed, \textit{``In the terms like in the way that it's spoken, and also kind of like the punctuation and stuff it helped it seem like a little bit more enthusiastic and friendly.''} (P1)

And once again, aligning with quantitative findings, P3 expressed pleasant surprise by \system’s warmth: \textit{``I can truthfully tell you that it was quite a pleasant experience because the algorithm that was being utilized really lent itself to empathy. ... I would feel incredibly willing and able to utilize this for longer stretches of time because the cordial and empathetic nature of [\system].``} (P3)

Finally, participants appreciated the realistic nature of the interactions with the system, noting that the conversations felt natural. They valued the system’s ability to engage in a back-and-forth dialogue that resembled genuine human conversation and its adaptability to the flow of discussion. As P3 shared, \textit{``It didn’t feel as if I was simply connecting the dots on behalf of an already pre-programmed algorithm or frameworks of conversational prompts. This one felt much more naturalistic, and as a result, if I had more to say or less to say, the AI would follow as naturally as I could, and I thought that was quite impressive.``} (P3)

These aspects of the system’s design were seen as key factors in its credibility and innovation, contributing positively to participants' experiences.

\section{Discussion}
\label{sec:discussion}

In this study, we examined people's general expectations of AI systems and compared how these expectations align with their evaluations of anthropomorphic versus non-anthropomorphic AI systems, using a repeated-measures design, incorporated both quantitative and qualitative methods. 

\subsection{Interpretation of Findings}
\label{sec:discussion-summary_of_findings}

Our findings reveal differences in participants’ expectations of AI systems compared to their actual evaluations of these systems, with slightly different patterns across the dimensions of relational warmth and perceived utility. 

Starting with relational warmth, the anthropomorphic AI system was rated as warmer than expected, whereas the non-anthropomorphic AI system did not differ in warmth from expectations. This suggests that anthropomorphic designs are particularly effective in conveying relational warmth. This finding is consistent with the qualitative feedback, where participants noted that the anthropomorphic system felt engaging and personable. 

On the other hand, when it came to perceived utility, both the anthropomorphic and non-anthropomorphic systems were rated as significantly more useful than participants expected, with no significant difference between the two. This suggests that anthropomorphic designs do not confer an additional boost to systems' perceived utility. Qualitative interviews supported these quantitative findings, with participants expressing pleasant surprise at the systems’ capabilities. 

While the anthropomorphic system was better at conveying warmth during interactions, both anthropomorphic and non-anthropomorphic systems were rated as more useful than participants initially expected. This suggests that the advantage of anthropomorphic design lies in enhancing relational warmth, but it does not necessarily provide additional benefits in terms of perceived utility.

Interestingly, when we examined behavioral engagement, no significant differences emerged between the anthropomorphic and non-anthropomorphic systems. Across both conditions, the majority of participants engaged with the recommended activities, pointing to the effectiveness of both systems in encouraging user engagement. However, this measure reflects engagement immediately after participants were first introduced to each system. Given that the anthropomorphic system was evaluated as being warmer than the non-anthropomorphic one, it remains possible that, over a longer period of use, the enhanced warmth of the anthropomorphic system could lead to higher rates of sustained engagement. This is an open question for future research. 

\subsection{Implications}
\label{sec:discussion-implications}

The unique advantage of anthropomorphic designs in conveying relational warmth (though not necessarily utility) suggests ways in which AI systems can be tailored to the context in which they are used. For example, in applications where building trust, empathy, or a sense of companionship is critical—such as mental health support—incorporating anthropomorphic elements could significantly enhance the user experience. In contrast, in contexts where utility and efficiency are the primary concerns, a more functional design may be sufficient, without the anthropomorphic component.

Our findings further suggest that people’s judgments of AI systems tend to revolve around two primary dimensions: relational warmth and perceived utility. This is consistent with seminal findings in social psychology that identify warmth and competence as fundamental dimensions of social perception \cite{fiske2007warmth}. Recent studies have shown that these dimensions also apply to how people perceive AI \cite{mckee2023warmth}. Although we refer to the factor in our study as "perceived utility," it aligns with the competence dimension observed in human judgments and reinforces the idea that the fundamental ways in which people evaluate other humans also extend to their perceptions of machines.

In addition, our findings contribute to the growing body of evidence suggesting a shift in how people perceive AI systems, particularly with the development of large language models and the increasing use of anthropomorphic designs. In earlier research, machines have been seen as low in “experience,” meaning they were perceived as incapable of emotions or forming emotional connections \cite{gray2007dimensions}. However, as LLMs improve and become more anthropomorphized, they increasingly facilitate interactions that convey more relational warmth. This aligns with prior research showing that anthropomorphizing AI systems can lead to higher perceptions of warmth \cite{kim2023empathy}. Our findings further suggest that anthropomorphic designs may not only meet, but \textit{exceed}, users’ expectations of warmth from AI systems.

Finally, another important implication of our findings is the gap between people's general expectations of AI systems and their actual experiences. On one hand, low expectations may discourage initial adoption, particularly in the context of well-being support. Yet, once users interact with these systems, positive evaluations show that firsthand experience can bridge this gap. On the other hand, research on expectation disconfirmation suggests that low expectations can lead to greater satisfaction when the system exceeds those initial assumptions~\cite{michalco2015relation}. Similarly, studies on AI metaphors show that projecting lower competence can result in more favorable evaluations when the AI outperforms expectations~\cite{khadpe2020metaphors}. Thus, while managing expectations is important, there may also be value in designing AI systems to pleasantly exceed user assumptions, turning underestimation into a tool for enhancing satisfaction.

\subsection{Limitations and Future Directions}
\label{sec:discussion-limitations}

This study has several limitations that warrant further investigation. First, the study was conducted over a relatively short period, spanning only a few days. Longer longitudinal studies are needed to assess whether differences in perceptions of AI systems ultimately lead to differences in continued engagement over time.

Additionally, we only tested one type of anthropomorphic design, specifically a sun character, rather than a human-like figure. This leaves open the question of whether different anthropomorphic designs might yield different effects on user perceptions and engagement. Future research should explore a wider variety of anthropomorphic designs to determine which features are most effective in enhancing user experience and fostering long-term engagement.

Additionally, the version of the AI system we tested was relatively limited in its level of personalization, since it recommended activities based on users' current mood. Expanding personalization to incorporate additional factors such as context (e.g., time, location), individual preferences (e.g., cognitive vs. physical activities), and ethical considerations (e.g., accommodations for disabilities) could further enhance its utility. For instance, it might not be appropriate to recommend a walk at midnight or to users with disabilities. Future research should explore how AI systems that account for these factors shape user perceptions.

Another limitation is that the sample size was relatively small and drawn from a Prolific sample based in the U.S., which limits the cultural generalizability of our findings. For instance, research in cultural psychology has found that values regarding social support and emotions differ across cultures, and these differences may moderate our findings. Specifically, while those in Western cultures tend to prefer explicit forms of support, those in East Asian cultures tend to prefer more implicit forms of support \cite{kim2008culture}. Similarly, in terms of the emotions people value, or ideal affect, European Americans tend to value high-arousal positive states (e.g., excitement) more than East Asians \cite{tsai2006cultural}. These cultural differences suggest that, all else being equal, perceptions of AI systems may vary across cultures, particularly if they are designed with Western values in mind.

Relatedly, our sample consisted of relatively healthy individuals, which may limit the applicability of our findings to other populations. It remains unclear how AI systems like the one we tested would be perceived by individuals with clinical conditions who may seek to use such systems as a complement to their weekly therapy sessions. Additionally, those with subclinical conditions might use the system proactively as a prevention tool. Future research should consider testing the system with more specific populations and psychological traits to better understand its effectiveness across different user groups.

Finally, an important area for future research is understanding how perceiving AI systems as warm might foster unintended emotional attachment. While anthropomorphizing technology can help simplify complex interactions and make AI more relatable, it may also lead users to form emotional dependencies or over-rely on the system \cite{darling2012whosjohnny}. Future studies should focus on balancing these pros and cons with the users’ best interests in mind, ensuring that warmth enhances user interaction without leading to unintended emotional consequences or manipulation.

\section{Conclusion}

In this paper, we examined the role of anthropomorphic and non-anthropomorphic AI systems in supporting users’ well-being by offering personalized, science-based activity recommendations. Our findings suggest that these AI systems can be surprisingly useful to users, and that anthropomorphic designs can further convey a surprising degree of relational warmth. 
Although our study focused on healthy populations, these findings raise the possibility that anthropomorphic AI systems could serve as supplementary tools in early-stage well-being support or as a bridge between therapy sessions. For instance, users might benefit from logging daily emotions and activities, offering valuable insights for both personal reflection and potential use by mental health professionals. Future research should explore these applications to better understand their potential role in supporting mental health.

\section*{Acknowledgement}

We acknowledge that the Intellectual Property of the prompt, the appearance, and persona design of the Sunnie system reported in this paper belong to Flourish Science Inc., a public benefit corporation. 
Figure~\ref{fig:teaser}
and~\ref{fig:sunnie_prompt} are licensed under a Creative Commons Attribution-NonCommercial-NoDerivs 4.0 International license. Copyright held by Flourish Science Inc. The user study was reviewed and approved by the Northeastern University IRB. We also thank all participants in our user study for their time and insights.

\bibliography{sample-base}


\begin{thebibliography}{139}


\ifx \showCODEN    \undefined \def \showCODEN     #1{\unskip}     \fi
\ifx \showDOI      \undefined \def \showDOI       #1{#1}\fi
\ifx \showISBNx    \undefined \def \showISBNx     #1{\unskip}     \fi
\ifx \showISBNxiii \undefined \def \showISBNxiii  #1{\unskip}     \fi
\ifx \showISSN     \undefined \def \showISSN      #1{\unskip}     \fi
\ifx \showLCCN     \undefined \def \showLCCN      #1{\unskip}     \fi
\ifx \shownote     \undefined \def \shownote      #1{#1}          \fi
\ifx \showarticletitle \undefined \def \showarticletitle #1{#1}   \fi
\ifx \showURL      \undefined \def \showURL       {\relax}        \fi
\providecommand\bibfield[2]{#2}
\providecommand\bibinfo[2]{#2}
\providecommand\natexlab[1]{#1}
\providecommand\showeprint[2][]{arXiv:#2}

\bibitem[Abd-Alrazaq et~al\mbox{.}(2020a)]%
        {Abd-Alrazaq2020}
\bibfield{author}{\bibinfo{person}{A.A. Abd-Alrazaq}, \bibinfo{person}{A. Rababeh}, \bibinfo{person}{M. Alajlani}, \bibinfo{person}{B.M. Bewick}, {and} \bibinfo{person}{M. Househ}.} \bibinfo{year}{2020}\natexlab{a}.
\newblock \showarticletitle{Effectiveness and Safety of Using Chatbots to Improve Mental Health: Systematic Review and Meta-Analysis}.
\newblock \bibinfo{journal}{\emph{Journal of Medical Internet Research}} \bibinfo{volume}{22}, \bibinfo{number}{7} (\bibinfo{year}{2020}), \bibinfo{pages}{e16021}.
\newblock
\urldef\tempurl%
\url{https://doi.org/10.2196/16021}
\showDOI{\tempurl}


\bibitem[Abd-Alrazaq et~al\mbox{.}(2020b)]%
        {abd2020effectiveness}
\bibfield{author}{\bibinfo{person}{Alaa~Ali Abd-Alrazaq}, \bibinfo{person}{Asma Rababeh}, \bibinfo{person}{Mohannad Alajlani}, \bibinfo{person}{Bridgette~M Bewick}, {and} \bibinfo{person}{Mowafa Househ}.} \bibinfo{year}{2020}\natexlab{b}.
\newblock \showarticletitle{Effectiveness and safety of using chatbots to improve mental health: systematic review and meta-analysis}.
\newblock \bibinfo{journal}{\emph{Journal of medical Internet research}} \bibinfo{volume}{22}, \bibinfo{number}{7} (\bibinfo{year}{2020}), \bibinfo{pages}{e16021}.
\newblock


\bibitem[Achiam et~al\mbox{.}(2023)]%
        {achiam2023gpt}
\bibfield{author}{\bibinfo{person}{Josh Achiam}, \bibinfo{person}{Steven Adler}, \bibinfo{person}{Sandhini Agarwal}, \bibinfo{person}{Lama Ahmad}, \bibinfo{person}{Ilge Akkaya}, \bibinfo{person}{Florencia~Leoni Aleman}, \bibinfo{person}{Diogo Almeida}, \bibinfo{person}{Janko Altenschmidt}, \bibinfo{person}{Sam Altman}, \bibinfo{person}{Shyamal Anadkat}, {et~al\mbox{.}}} \bibinfo{year}{2023}\natexlab{}.
\newblock \showarticletitle{Gpt-4 technical report}.
\newblock \bibinfo{journal}{\emph{arXiv preprint arXiv:2303.08774}} (\bibinfo{year}{2023}).
\newblock


\bibitem[Aggarwal et~al\mbox{.}(2023)]%
        {Aggarwal_Tam_Wu_Li_Qiao_2023}
\bibfield{author}{\bibinfo{person}{Abhishek Aggarwal}, \bibinfo{person}{Cheuk~Chi Tam}, \bibinfo{person}{Dezhi Wu}, \bibinfo{person}{Xiaoming Li}, {and} \bibinfo{person}{Shan Qiao}.} \bibinfo{year}{2023}\natexlab{}.
\newblock \showarticletitle{Artificial Intelligence–Based Chatbots for Promoting Health Behavioral Changes: Systematic Review}.
\newblock \bibinfo{journal}{\emph{Journal of Medical Internet Research}} \bibinfo{volume}{25}, \bibinfo{number}{1} (\bibinfo{date}{Feb.} \bibinfo{year}{2023}), \bibinfo{pages}{e40789}.
\newblock
\urldef\tempurl%
\url{https://doi.org/10.2196/40789}
\showDOI{\tempurl}


\bibitem[Aron et~al\mbox{.}(1997)]%
        {aron1997experimental}
\bibfield{author}{\bibinfo{person}{Arthur Aron}, \bibinfo{person}{Edward Melinat}, \bibinfo{person}{Elaine~N Aron}, \bibinfo{person}{Robert~Darrin Vallone}, {and} \bibinfo{person}{Renee~J Bator}.} \bibinfo{year}{1997}\natexlab{}.
\newblock \showarticletitle{The experimental generation of interpersonal closeness: A procedure and some preliminary findings}.
\newblock \bibinfo{journal}{\emph{Personality and social psychology bulletin}} \bibinfo{volume}{23}, \bibinfo{number}{4} (\bibinfo{year}{1997}), \bibinfo{pages}{363--377}.
\newblock


\bibitem[Association(2022)]%
        {apa2022impact}
\bibfield{author}{\bibinfo{person}{American~Psychological Association}.} \bibinfo{year}{2022}\natexlab{}.
\newblock \bibinfo{title}{Psychologists struggle to meet demand amid mental health crisis: 2022 COVID-19 Practitioner Impact Survey}.
\newblock
\newblock
\urldef\tempurl%
\url{https://www.apa.org/news/press/releases/2022/11/mental-health-crisis-demand}
\showURL{%
\tempurl}
\newblock
\shownote{Accessed: 2023-09-10}.


\bibitem[Bickmore and Cassell(2001)]%
        {bickmore2001relational}
\bibfield{author}{\bibinfo{person}{Timothy Bickmore} {and} \bibinfo{person}{Justine Cassell}.} \bibinfo{year}{2001}\natexlab{}.
\newblock \showarticletitle{Relational agents: a model and implementation of building user trust}. In \bibinfo{booktitle}{\emph{Proceedings of the SIGCHI conference on Human factors in computing systems}}. \bibinfo{pages}{396--403}.
\newblock


\bibitem[Bickmore and Ring(2010)]%
        {bickmore2010making}
\bibfield{author}{\bibinfo{person}{Timothy Bickmore} {and} \bibinfo{person}{Lazlo Ring}.} \bibinfo{year}{2010}\natexlab{}.
\newblock \showarticletitle{Making it personal: end-user authoring of health narratives delivered by virtual agents}. In \bibinfo{booktitle}{\emph{Intelligent Virtual Agents: 10th International Conference, IVA 2010, Philadelphia, PA, USA, September 20-22, 2010. Proceedings 10}}. Springer, \bibinfo{pages}{399--405}.
\newblock


\bibitem[Bickmore et~al\mbox{.}(2005)]%
        {bickmore2005s}
\bibfield{author}{\bibinfo{person}{Timothy~W Bickmore}, \bibinfo{person}{Lisa Caruso}, \bibinfo{person}{Kerri Clough-Gorr}, {and} \bibinfo{person}{Tim Heeren}.} \bibinfo{year}{2005}\natexlab{}.
\newblock \showarticletitle{‘It’s just like you talk to a friend’relational agents for older adults}.
\newblock \bibinfo{journal}{\emph{Interacting with Computers}} \bibinfo{volume}{17}, \bibinfo{number}{6} (\bibinfo{year}{2005}), \bibinfo{pages}{711--735}.
\newblock


\bibitem[Bickmore and Picard(2005)]%
        {bickmore2005establishing}
\bibfield{author}{\bibinfo{person}{Timothy~W Bickmore} {and} \bibinfo{person}{Rosalind~W Picard}.} \bibinfo{year}{2005}\natexlab{}.
\newblock \showarticletitle{Establishing and maintaining long-term human-computer relationships}.
\newblock \bibinfo{journal}{\emph{ACM Transactions on Computer-Human Interaction (TOCHI)}} \bibinfo{volume}{12}, \bibinfo{number}{2} (\bibinfo{year}{2005}), \bibinfo{pages}{293--327}.
\newblock


\bibitem[Bijkerk et~al\mbox{.}(2023)]%
        {bijkerk2023measuring}
\bibfield{author}{\bibinfo{person}{Laura~Esther Bijkerk}, \bibinfo{person}{Anke Oenema}, \bibinfo{person}{Nicole Geschwind}, {and} \bibinfo{person}{Mark Spigt}.} \bibinfo{year}{2023}\natexlab{}.
\newblock \showarticletitle{Measuring engagement with mental health and behavior change interventions: an integrative review of methods and instruments}.
\newblock \bibinfo{journal}{\emph{International Journal of Behavioral Medicine}} \bibinfo{volume}{30}, \bibinfo{number}{2} (\bibinfo{year}{2023}), \bibinfo{pages}{155--166}.
\newblock


\bibitem[Bluvstein et~al\mbox{.}(2024)]%
        {bluvstein2024imperfectly}
\bibfield{author}{\bibinfo{person}{Shirley Bluvstein}, \bibinfo{person}{Xuan Zhao}, \bibinfo{person}{Alixandra Barasch}, {and} \bibinfo{person}{Juliana Schroeder}.} \bibinfo{year}{2024}\natexlab{}.
\newblock \showarticletitle{Imperfectly Human: The Humanizing Potential of (Corrected) Errors in Text-Based Communication}.
\newblock \bibinfo{journal}{\emph{Journal of the Association for Consumer Research}} \bibinfo{volume}{9}, \bibinfo{number}{3} (\bibinfo{year}{2024}), \bibinfo{pages}{000--000}.
\newblock


\bibitem[Boduszek et~al\mbox{.}(2019)]%
        {boduszek2019prosocial}
\bibfield{author}{\bibinfo{person}{Daniel Boduszek}, \bibinfo{person}{Agata Debowska}, \bibinfo{person}{Adele~D Jones}, \bibinfo{person}{Minhua Ma}, \bibinfo{person}{David Smith}, \bibinfo{person}{Dominic Willmott}, \bibinfo{person}{Ena~Trotman Jemmott}, \bibinfo{person}{Hazel Da~Breo}, {and} \bibinfo{person}{Gillian Kirkman}.} \bibinfo{year}{2019}\natexlab{}.
\newblock \showarticletitle{Prosocial video game as an intimate partner violence prevention tool among youth: A randomised controlled trial}.
\newblock \bibinfo{journal}{\emph{Computers in Human Behavior}}  \bibinfo{volume}{93} (\bibinfo{year}{2019}), \bibinfo{pages}{260--266}.
\newblock


\bibitem[Bowman et~al\mbox{.}(2024)]%
        {Bowman_Cooney_Newbold_Thieme_Clark_Doherty_Cowan_2024}
\bibfield{author}{\bibinfo{person}{Robert Bowman}, \bibinfo{person}{Orla Cooney}, \bibinfo{person}{Joseph~W. Newbold}, \bibinfo{person}{Anja Thieme}, \bibinfo{person}{Leigh Clark}, \bibinfo{person}{Gavin Doherty}, {and} \bibinfo{person}{Benjamin Cowan}.} \bibinfo{year}{2024}\natexlab{}.
\newblock \showarticletitle{Exploring how politeness impacts the user experience of chatbots for mental health support}.
\newblock \bibinfo{journal}{\emph{International Journal of Human-Computer Studies}}  \bibinfo{volume}{184} (\bibinfo{year}{2024}), \bibinfo{pages}{103181}.
\newblock
\showISSN{1071-5819}
\urldef\tempurl%
\url{https://doi.org/10.1016/j.ijhcs.2023.103181}
\showDOI{\tempurl}


\bibitem[Chan et~al\mbox{.}(2023)]%
        {chan2023mango}
\bibfield{author}{\bibinfo{person}{Szeyi Chan}, \bibinfo{person}{Jiachen Li}, \bibinfo{person}{Bingsheng Yao}, \bibinfo{person}{Amama Mahmood}, \bibinfo{person}{Chien-Ming Huang}, \bibinfo{person}{Holly Jimison}, \bibinfo{person}{Elizabeth~D Mynatt}, {and} \bibinfo{person}{Dakuo Wang}.} \bibinfo{year}{2023}\natexlab{}.
\newblock \showarticletitle{" Mango Mango, How to Let The Lettuce Dry Without A Spinner?'': Exploring User Perceptions of Using An LLM-Based Conversational Assistant Toward Cooking Partner}.
\newblock \bibinfo{journal}{\emph{arXiv preprint arXiv:2310.05853}} (\bibinfo{year}{2023}).
\newblock


\bibitem[Chinmulgund et~al\mbox{.}(2023)]%
        {chinmulgund2023anthropomorphism}
\bibfield{author}{\bibinfo{person}{Avanti Chinmulgund}, \bibinfo{person}{Ritesh Khatwani}, \bibinfo{person}{Poornima Tapas}, \bibinfo{person}{Pritesh Shah}, {and} \bibinfo{person}{Ravi Sekhar}.} \bibinfo{year}{2023}\natexlab{}.
\newblock \showarticletitle{Anthropomorphism of AI based chatbots by users during communication}. In \bibinfo{booktitle}{\emph{2023 3rd International Conference on Intelligent Technologies (CONIT)}}. \bibinfo{publisher}{IEEE}, \bibinfo{pages}{1--6}.
\newblock
\urldef\tempurl%
\url{https://doi.org/10.1109/CONIT59222.2023.10205689}
\showDOI{\tempurl}


\bibitem[Chishima et~al\mbox{.}(2021)]%
        {chishima2021temporal}
\bibfield{author}{\bibinfo{person}{Yuta Chishima}, \bibinfo{person}{I-Ting Huai-Ching~Liu}, {and} \bibinfo{person}{Anne E.~Wilson}.} \bibinfo{year}{2021}\natexlab{}.
\newblock \showarticletitle{Temporal distancing during the COVID-19 pandemic: Letter writing with future self can mitigate negative affect}.
\newblock \bibinfo{journal}{\emph{Applied Psychology: Health and Well-Being}} \bibinfo{volume}{13}, \bibinfo{number}{2} (\bibinfo{year}{2021}), \bibinfo{pages}{406--418}.
\newblock


\bibitem[Cho et~al\mbox{.}(2019)]%
        {cho2019once}
\bibfield{author}{\bibinfo{person}{Min~Kyung Cho}, \bibinfo{person}{Seung~Jun Lee}, {and} \bibinfo{person}{Kun-Pyo Lee}.} \bibinfo{year}{2019}\natexlab{}.
\newblock \showarticletitle{Once a kind friend is now a thing: Understanding how conversational agents at home are forgotten}. In \bibinfo{booktitle}{\emph{Proceedings of the 2019 on designing interactive systems conference}}. \bibinfo{pages}{1557--1569}.
\newblock


\bibitem[Cho et~al\mbox{.}(2023)]%
        {cho2023evaluating}
\bibfield{author}{\bibinfo{person}{Yujin Cho}, \bibinfo{person}{Mingeon Kim}, \bibinfo{person}{Seojin Kim}, \bibinfo{person}{Oyun Kwon}, \bibinfo{person}{Ryan~Donghan Kwon}, \bibinfo{person}{Yoonha Lee}, {and} \bibinfo{person}{Dohyun Lim}.} \bibinfo{year}{2023}\natexlab{}.
\newblock \bibinfo{title}{Evaluating the Efficacy of Interactive Language Therapy Based on LLM for High-Functioning Autistic Adolescent Psychological Counseling}.
\newblock
\newblock
\showeprint[arxiv]{2311.09243}~[cs.HC]


\bibitem[Clark et~al\mbox{.}(2019)]%
        {clark2019makes}
\bibfield{author}{\bibinfo{person}{Leigh Clark}, \bibinfo{person}{Nadia Pantidi}, \bibinfo{person}{Orla Cooney}, \bibinfo{person}{Philip Doyle}, \bibinfo{person}{Diego Garaialde}, \bibinfo{person}{Justin Edwards}, \bibinfo{person}{Brendan Spillane}, \bibinfo{person}{Emer Gilmartin}, \bibinfo{person}{Christine Murad}, \bibinfo{person}{Cosmin Munteanu}, {et~al\mbox{.}}} \bibinfo{year}{2019}\natexlab{}.
\newblock \showarticletitle{What makes a good conversation? Challenges in designing truly conversational agents}. In \bibinfo{booktitle}{\emph{Proceedings of the 2019 CHI conference on human factors in computing systems}}. \bibinfo{pages}{1--12}.
\newblock


\bibitem[Consolvo et~al\mbox{.}(2009)]%
        {consolvo2009theory}
\bibfield{author}{\bibinfo{person}{Sunny Consolvo}, \bibinfo{person}{David~W McDonald}, {and} \bibinfo{person}{James~A Landay}.} \bibinfo{year}{2009}\natexlab{}.
\newblock \showarticletitle{Theory-driven design strategies for technologies that support behavior change in everyday life}. In \bibinfo{booktitle}{\emph{Proceedings of the SIGCHI conference on human factors in computing systems}}. \bibinfo{pages}{405--414}.
\newblock


\bibitem[Darling(2012)]%
        {darling2012whosjohnny}
\bibfield{author}{\bibinfo{person}{Kate Darling}.} \bibinfo{year}{2012}\natexlab{}.
\newblock \showarticletitle{Who's Johnny? Anthropomorphic Framing in Human-Robot Interaction, Integration, and Policy}.
\newblock \bibinfo{journal}{\emph{Proceedings of We Robot Conference}} (\bibinfo{year}{2012}).
\newblock


\bibitem[De~Choudhury et~al\mbox{.}(2023)]%
        {de2023benefits}
\bibfield{author}{\bibinfo{person}{Munmun De~Choudhury}, \bibinfo{person}{Sachin~R Pendse}, {and} \bibinfo{person}{Neha Kumar}.} \bibinfo{year}{2023}\natexlab{}.
\newblock \showarticletitle{Benefits and harms of large language models in digital mental health}.
\newblock \bibinfo{journal}{\emph{arXiv preprint arXiv:2311.14693}} (\bibinfo{year}{2023}).
\newblock


\bibitem[De~Visser et~al\mbox{.}(2016)]%
        {de2016almost}
\bibfield{author}{\bibinfo{person}{Ewart~J De~Visser}, \bibinfo{person}{Samuel~S Monfort}, \bibinfo{person}{Ryan McKendrick}, \bibinfo{person}{Melissa~AB Smith}, \bibinfo{person}{Patrick~E McKnight}, \bibinfo{person}{Frank Krueger}, {and} \bibinfo{person}{Raja Parasuraman}.} \bibinfo{year}{2016}\natexlab{}.
\newblock \showarticletitle{Almost human: Anthropomorphism increases trust resilience in cognitive agents.}
\newblock \bibinfo{journal}{\emph{Journal of Experimental Psychology: Applied}} \bibinfo{volume}{22}, \bibinfo{number}{3} (\bibinfo{year}{2016}), \bibinfo{pages}{331}.
\newblock


\bibitem[Denecke et~al\mbox{.}(2021)]%
        {Denecke21}
\bibfield{author}{\bibinfo{person}{Kerstin Denecke}, \bibinfo{person}{Sayan Vaaheesan}, {and} \bibinfo{person}{Aaganya Arulnathan}.} \bibinfo{year}{2021}\natexlab{}.
\newblock \showarticletitle{A Mental Health Chatbot for Regulating Emotions (SERMO) - Concept and Usability Test}.
\newblock \bibinfo{journal}{\emph{IEEE Transactions on Emerging Topics in Computing}} \bibinfo{volume}{9}, \bibinfo{number}{3} (\bibinfo{year}{2021}), \bibinfo{pages}{1170--1182}.
\newblock
\urldef\tempurl%
\url{https://doi.org/10.1109/TETC.2020.2974478}
\showDOI{\tempurl}


\bibitem[Dweck(2006)]%
        {dweck2006mindset}
\bibfield{author}{\bibinfo{person}{Carol~S Dweck}.} \bibinfo{year}{2006}\natexlab{}.
\newblock \bibinfo{booktitle}{\emph{Mindset: The new psychology of success}}.
\newblock \bibinfo{publisher}{Random house}.
\newblock


\bibitem[Epley et~al\mbox{.}(2022)]%
        {epley2022undersociality}
\bibfield{author}{\bibinfo{person}{Nicholas Epley}, \bibinfo{person}{Michael Kardas}, \bibinfo{person}{Xuan Zhao}, \bibinfo{person}{Stav Atir}, {and} \bibinfo{person}{Juliana Schroeder}.} \bibinfo{year}{2022}\natexlab{}.
\newblock \showarticletitle{Undersociality: Miscalibrated social cognition can inhibit social connection}.
\newblock \bibinfo{journal}{\emph{Trends in Cognitive Sciences}} \bibinfo{volume}{26}, \bibinfo{number}{5} (\bibinfo{year}{2022}), \bibinfo{pages}{406--418}.
\newblock


\bibitem[Fadhil et~al\mbox{.}(2019)]%
        {fadhil2019assistive}
\bibfield{author}{\bibinfo{person}{Ahmed Fadhil}, \bibinfo{person}{Yunlong Wang}, {and} \bibinfo{person}{Harald Reiterer}.} \bibinfo{year}{2019}\natexlab{}.
\newblock \showarticletitle{Assistive conversational agent for health coaching: a validation study}.
\newblock \bibinfo{journal}{\emph{Methods of information in medicine}} \bibinfo{volume}{58}, \bibinfo{number}{01} (\bibinfo{year}{2019}), \bibinfo{pages}{009--023}.
\newblock


\bibitem[Fan et~al\mbox{.}(2021)]%
        {fan2021utilization}
\bibfield{author}{\bibinfo{person}{Xiangmin Fan}, \bibinfo{person}{Daren Chao}, \bibinfo{person}{Zhan Zhang}, \bibinfo{person}{Dakuo Wang}, \bibinfo{person}{Xiaohua Li}, {and} \bibinfo{person}{Feng Tian}.} \bibinfo{year}{2021}\natexlab{}.
\newblock \showarticletitle{Utilization of self-diagnosis health chatbots in real-world settings: case study}.
\newblock \bibinfo{journal}{\emph{Journal of medical Internet research}} \bibinfo{volume}{23}, \bibinfo{number}{1} (\bibinfo{year}{2021}), \bibinfo{pages}{e19928}.
\newblock


\bibitem[Feine et~al\mbox{.}(2019)]%
        {feine2019taxonomy}
\bibfield{author}{\bibinfo{person}{Jasper Feine}, \bibinfo{person}{Ulrich Gnewuch}, \bibinfo{person}{Stefan Morana}, {and} \bibinfo{person}{Alexander Maedche}.} \bibinfo{year}{2019}\natexlab{}.
\newblock \showarticletitle{A taxonomy of social cues for conversational agents}.
\newblock \bibinfo{journal}{\emph{International Journal of Human-Computer Studies}}  \bibinfo{volume}{132} (\bibinfo{year}{2019}), \bibinfo{pages}{138--161}.
\newblock


\bibitem[Fiske et~al\mbox{.}(2007)]%
        {fiske2007warmth}
\bibfield{author}{\bibinfo{person}{Susan~T Fiske}, \bibinfo{person}{Amy~JC Cuddy}, {and} \bibinfo{person}{Peter Glick}.} \bibinfo{year}{2007}\natexlab{}.
\newblock \showarticletitle{Universal dimensions of social cognition: Warmth and competence}.
\newblock \bibinfo{journal}{\emph{Trends in Cognitive Sciences}} \bibinfo{volume}{11}, \bibinfo{number}{2} (\bibinfo{year}{2007}), \bibinfo{pages}{77--83}.
\newblock


\bibitem[Fitzpatrick et~al\mbox{.}(2017)]%
        {Fitzpatrick2017}
\bibfield{author}{\bibinfo{person}{Kathleen~Kara Fitzpatrick}, \bibinfo{person}{Alison Darcy}, {and} \bibinfo{person}{Molly Vierhile}.} \bibinfo{year}{2017}\natexlab{}.
\newblock \showarticletitle{Delivering Cognitive Behavior Therapy to Young Adults With Symptoms of Depression and Anxiety Using a Fully Automated Conversational Agent (Woebot): A Randomized Controlled Trial}.
\newblock \bibinfo{journal}{\emph{JMIR Mental Health}} \bibinfo{volume}{4}, \bibinfo{number}{2} (\bibinfo{year}{2017}), \bibinfo{pages}{e19}.
\newblock
\urldef\tempurl%
\url{https://doi.org/10.2196/mental.7785}
\showDOI{\tempurl}


\bibitem[Fredrickson(2001)]%
        {fredrickson2001role}
\bibfield{author}{\bibinfo{person}{Barbara~L Fredrickson}.} \bibinfo{year}{2001}\natexlab{}.
\newblock \showarticletitle{The role of positive emotions in positive psychology: The broaden-and-build theory of positive emotions.}
\newblock \bibinfo{journal}{\emph{American psychologist}} \bibinfo{volume}{56}, \bibinfo{number}{3} (\bibinfo{year}{2001}), \bibinfo{pages}{218}.
\newblock


\bibitem[Gable et~al\mbox{.}(2018)]%
        {gable2018you}
\bibfield{author}{\bibinfo{person}{Shelly~L Gable}, \bibinfo{person}{Harry~T Reis}, \bibinfo{person}{Emily~A Impett}, {and} \bibinfo{person}{Evan~R Asher}.} \bibinfo{year}{2018}\natexlab{}.
\newblock \showarticletitle{What do you do when things go right? The intrapersonal and interpersonal benefits of sharing positive events}.
\newblock In \bibinfo{booktitle}{\emph{Relationships, well-being and behaviour}}. \bibinfo{publisher}{Routledge}, \bibinfo{pages}{144--182}.
\newblock


\bibitem[Gnewuch et~al\mbox{.}(2018)]%
        {gnewuch2018faster}
\bibfield{author}{\bibinfo{person}{Ulrich Gnewuch}, \bibinfo{person}{Stefan Morana}, \bibinfo{person}{Marc Adam}, {and} \bibinfo{person}{Alexander Maedche}.} \bibinfo{year}{2018}\natexlab{}.
\newblock \showarticletitle{Faster is not always better: understanding the effect of dynamic response delays in human-chatbot interaction}.
\newblock  (\bibinfo{year}{2018}).
\newblock


\bibitem[Gnewuch et~al\mbox{.}(2017)]%
        {gnewuch2017towards}
\bibfield{author}{\bibinfo{person}{Ulrich Gnewuch}, \bibinfo{person}{Stefan Morana}, {and} \bibinfo{person}{Alexander Maedche}.} \bibinfo{year}{2017}\natexlab{}.
\newblock \showarticletitle{Towards Designing Cooperative and Social Conversational Agents for Customer Service.}. In \bibinfo{booktitle}{\emph{ICIS}}. \bibinfo{pages}{1--13}.
\newblock


\bibitem[Gold et~al\mbox{.}(2023)]%
        {gold2023three}
\bibfield{author}{\bibinfo{person}{Katherine~J Gold}, \bibinfo{person}{Margaret~L Dobson}, {and} \bibinfo{person}{Ananda Sen}.} \bibinfo{year}{2023}\natexlab{}.
\newblock \showarticletitle{“Three Good Things” Digital Intervention Among Health Care Workers: A Randomized Controlled Trial}.
\newblock \bibinfo{journal}{\emph{The Annals of Family Medicine}} \bibinfo{volume}{21}, \bibinfo{number}{3} (\bibinfo{year}{2023}), \bibinfo{pages}{220--226}.
\newblock


\bibitem[Grassini(2022)]%
        {grassini2022systematic}
\bibfield{author}{\bibinfo{person}{Simone Grassini}.} \bibinfo{year}{2022}\natexlab{}.
\newblock \showarticletitle{A systematic review and meta-analysis of nature walk as an intervention for anxiety and depression}.
\newblock \bibinfo{journal}{\emph{Journal of clinical medicine}} \bibinfo{volume}{11}, \bibinfo{number}{6} (\bibinfo{year}{2022}), \bibinfo{pages}{1731}.
\newblock


\bibitem[Gray et~al\mbox{.}(2007)]%
        {gray2007dimensions}
\bibfield{author}{\bibinfo{person}{Kurt Gray}, \bibinfo{person}{Heather~M Gray}, {and} \bibinfo{person}{Daniel~M Wegner}.} \bibinfo{year}{2007}\natexlab{}.
\newblock \showarticletitle{Dimensions of mind perception}.
\newblock \bibinfo{journal}{\emph{Science}} \bibinfo{volume}{315}, \bibinfo{number}{5812} (\bibinfo{year}{2007}), \bibinfo{pages}{619}.
\newblock


\bibitem[Grov{\'e}(2021)]%
        {grove2021co}
\bibfield{author}{\bibinfo{person}{Christine Grov{\'e}}.} \bibinfo{year}{2021}\natexlab{}.
\newblock \showarticletitle{Co-developing a mental health and wellbeing chatbot with and for young people}.
\newblock \bibinfo{journal}{\emph{Frontiers in psychiatry}}  \bibinfo{volume}{11} (\bibinfo{year}{2021}), \bibinfo{pages}{606041}.
\newblock


\bibitem[Ho et~al\mbox{.}(2018)]%
        {ho2018psychological}
\bibfield{author}{\bibinfo{person}{Annabell Ho}, \bibinfo{person}{Jeff Hancock}, {and} \bibinfo{person}{Adam~S Miner}.} \bibinfo{year}{2018}\natexlab{}.
\newblock \showarticletitle{Psychological, relational, and emotional effects of self-disclosure after conversations with a chatbot}.
\newblock \bibinfo{journal}{\emph{Journal of Communication}} \bibinfo{volume}{68}, \bibinfo{number}{4} (\bibinfo{year}{2018}), \bibinfo{pages}{712--733}.
\newblock


\bibitem[Hua et~al\mbox{.}(2024)]%
        {Hua2024LargeLM}
\bibfield{author}{\bibinfo{person}{Yining Hua}, \bibinfo{person}{Fenglin Liu}, \bibinfo{person}{Kailai Yang}, \bibinfo{person}{Zehan Li}, \bibinfo{person}{Yi han Sheu}, \bibinfo{person}{Peilin Zhou}, \bibinfo{person}{Lauren~V. Moran}, \bibinfo{person}{Sophia Ananiadou}, {and} \bibinfo{person}{Andrew Beam}.} \bibinfo{year}{2024}\natexlab{}.
\newblock \showarticletitle{Large Language Models in Mental Health Care: a Scoping Review}.
\newblock \bibinfo{journal}{\emph{ArXiv}}  \bibinfo{volume}{abs/2401.02984} (\bibinfo{year}{2024}).
\newblock
\urldef\tempurl%
\url{https://api.semanticscholar.org/CorpusID:266843868}
\showURL{%
\tempurl}


\bibitem[Inkster et~al\mbox{.}(2018a)]%
        {Inkster2018AnEC}
\bibfield{author}{\bibinfo{person}{Becky Inkster}, \bibinfo{person}{Shubhankar Sarda}, {and} \bibinfo{person}{Vinod Subramanian}.} \bibinfo{year}{2018}\natexlab{a}.
\newblock \showarticletitle{An Empathy-Driven, Conversational Artificial Intelligence Agent (Wysa) for Digital Mental Well-Being: Real-World Data Evaluation Mixed-Methods Study}.
\newblock \bibinfo{journal}{\emph{JMIR mHealth and uHealth}}  \bibinfo{volume}{6} (\bibinfo{year}{2018}).
\newblock
\urldef\tempurl%
\url{https://api.semanticscholar.org/CorpusID:53719693}
\showURL{%
\tempurl}


\bibitem[Inkster et~al\mbox{.}(2018b)]%
        {inkster2018empathy}
\bibfield{author}{\bibinfo{person}{Becky Inkster}, \bibinfo{person}{Shubhankar Sarda}, \bibinfo{person}{Vinod Subramanian}, {et~al\mbox{.}}} \bibinfo{year}{2018}\natexlab{b}.
\newblock \showarticletitle{An empathy-driven, conversational artificial intelligence agent (Wysa) for digital mental well-being: real-world data evaluation mixed-methods study}.
\newblock \bibinfo{journal}{\emph{JMIR mHealth and uHealth}} \bibinfo{volume}{6}, \bibinfo{number}{11} (\bibinfo{year}{2018}), \bibinfo{pages}{e12106}.
\newblock


\bibitem[Isbister and Nass(2000)]%
        {isbister2000consistency}
\bibfield{author}{\bibinfo{person}{Katherine Isbister} {and} \bibinfo{person}{Clifford Nass}.} \bibinfo{year}{2000}\natexlab{}.
\newblock \showarticletitle{Consistency of personality in interactive characters: verbal cues, non-verbal cues, and user characteristics}.
\newblock \bibinfo{journal}{\emph{International journal of human-computer studies}} \bibinfo{volume}{53}, \bibinfo{number}{2} (\bibinfo{year}{2000}), \bibinfo{pages}{251--267}.
\newblock


\bibitem[Jorm et~al\mbox{.}(1997)]%
        {jorm1997mental}
\bibfield{author}{\bibinfo{person}{Anthony~F Jorm}, \bibinfo{person}{Ailsa~E Korten}, \bibinfo{person}{Patricia~A Jacomb}, \bibinfo{person}{Helen Christensen}, \bibinfo{person}{Bryan Rodgers}, {and} \bibinfo{person}{Penelope Pollitt}.} \bibinfo{year}{1997}\natexlab{}.
\newblock \showarticletitle{“Mental health literacy”: a survey of the public's ability to recognise mental disorders and their beliefs about the effectiveness of treatment}.
\newblock \bibinfo{journal}{\emph{Medical journal of Australia}} \bibinfo{volume}{166}, \bibinfo{number}{4} (\bibinfo{year}{1997}), \bibinfo{pages}{182--186}.
\newblock


\bibitem[Jose et~al\mbox{.}(2012)]%
        {jose2012does}
\bibfield{author}{\bibinfo{person}{Paul~E Jose}, \bibinfo{person}{Bee~T Lim}, {and} \bibinfo{person}{Fred~B Bryant}.} \bibinfo{year}{2012}\natexlab{}.
\newblock \showarticletitle{Does savoring increase happiness? A daily diary study}.
\newblock \bibinfo{journal}{\emph{The Journal of Positive Psychology}} \bibinfo{volume}{7}, \bibinfo{number}{3} (\bibinfo{year}{2012}), \bibinfo{pages}{176--187}.
\newblock


\bibitem[Kabat-Zinn(2003)]%
        {kabat2003mindfulness}
\bibfield{author}{\bibinfo{person}{Jon Kabat-Zinn}.} \bibinfo{year}{2003}\natexlab{}.
\newblock \showarticletitle{Mindfulness-based interventions in context: past, present, and future.}
\newblock  (\bibinfo{year}{2003}).
\newblock


\bibitem[Kardas et~al\mbox{.}(2022)]%
        {kardas2022overly}
\bibfield{author}{\bibinfo{person}{Michael Kardas}, \bibinfo{person}{Amit Kumar}, {and} \bibinfo{person}{Nicholas Epley}.} \bibinfo{year}{2022}\natexlab{}.
\newblock \showarticletitle{Overly shallow?: Miscalibrated expectations create a barrier to deeper conversation.}
\newblock \bibinfo{journal}{\emph{Journal of Personality and Social Psychology}} \bibinfo{volume}{122}, \bibinfo{number}{3} (\bibinfo{year}{2022}), \bibinfo{pages}{367}.
\newblock


\bibitem[Keyes(2002)]%
        {Keyes2002}
\bibfield{author}{\bibinfo{person}{Corey L.~M. Keyes}.} \bibinfo{year}{2002}\natexlab{}.
\newblock \showarticletitle{The mental health continuum: from languishing to flourishing in life}.
\newblock \bibinfo{journal}{\emph{Journal of Health and Social Behavior}} \bibinfo{volume}{43}, \bibinfo{number}{2} (\bibinfo{date}{Jun} \bibinfo{year}{2002}), \bibinfo{pages}{207--222}.
\newblock


\bibitem[Khadpe et~al\mbox{.}(2020)]%
        {khadpe2020metaphors}
\bibfield{author}{\bibinfo{person}{Pranav Khadpe}, \bibinfo{person}{Ranjay Krishna}, \bibinfo{person}{Li Fei-Fei}, \bibinfo{person}{Jeff~T Hancock}, {and} \bibinfo{person}{Michael~S Bernstein}.} \bibinfo{year}{2020}\natexlab{}.
\newblock \showarticletitle{Conceptual metaphors impact perceptions of human-ai collaboration}.
\newblock \bibinfo{journal}{\emph{Proceedings of the ACM on Human-Computer Interaction}} \bibinfo{volume}{4}, \bibinfo{number}{CSCW2} (\bibinfo{year}{2020}), \bibinfo{pages}{1--26}.
\newblock


\bibitem[Kim et~al\mbox{.}(2008)]%
        {kim2008culture}
\bibfield{author}{\bibinfo{person}{Heejung~S. Kim}, \bibinfo{person}{David~K. Sherman}, {and} \bibinfo{person}{Shelley~E. Taylor}.} \bibinfo{year}{2008}\natexlab{}.
\newblock \showarticletitle{Culture and social support}.
\newblock \bibinfo{journal}{\emph{American Psychologist}} \bibinfo{volume}{63}, \bibinfo{number}{6} (\bibinfo{year}{2008}), \bibinfo{pages}{518--526}.
\newblock


\bibitem[Kim and Hur(2023)]%
        {kim2023empathy}
\bibfield{author}{\bibinfo{person}{Woo~Bin Kim} {and} \bibinfo{person}{Hyun~Jung Hur}.} \bibinfo{year}{2023}\natexlab{}.
\newblock \showarticletitle{What Makes People Feel Empathy for AI Chatbots? Assessing the Role of Competence and Warmth}.
\newblock \bibinfo{journal}{\emph{International Journal of Human--Computer Interaction}} \bibinfo{volume}{40}, \bibinfo{number}{17} (\bibinfo{year}{2023}), \bibinfo{pages}{4674--4687}.
\newblock
\urldef\tempurl%
\url{https://doi.org/10.1080/10447318.2023.2219961}
\showDOI{\tempurl}


\bibitem[Knittle et~al\mbox{.}(2018)]%
        {knittle2018can}
\bibfield{author}{\bibinfo{person}{Keegan Knittle}, \bibinfo{person}{Johanna Nurmi}, \bibinfo{person}{Rik Crutzen}, \bibinfo{person}{Nelli Hankonen}, \bibinfo{person}{Marguerite Beattie}, {and} \bibinfo{person}{Stephan~U Dombrowski}.} \bibinfo{year}{2018}\natexlab{}.
\newblock \showarticletitle{How can interventions increase motivation for physical activity? A systematic review and meta-analysis}.
\newblock \bibinfo{journal}{\emph{Health psychology review}} \bibinfo{volume}{12}, \bibinfo{number}{3} (\bibinfo{year}{2018}), \bibinfo{pages}{211--230}.
\newblock


\bibitem[Kocielnik et~al\mbox{.}(2018)]%
        {kocielnik2018reflection}
\bibfield{author}{\bibinfo{person}{Rafal Kocielnik}, \bibinfo{person}{Lillian Xiao}, \bibinfo{person}{Daniel Avrahami}, {and} \bibinfo{person}{Gary Hsieh}.} \bibinfo{year}{2018}\natexlab{}.
\newblock \showarticletitle{Reflection companion: a conversational system for engaging users in reflection on physical activity}.
\newblock \bibinfo{journal}{\emph{Proceedings of the ACM on Interactive, Mobile, Wearable and Ubiquitous Technologies}} \bibinfo{volume}{2}, \bibinfo{number}{2} (\bibinfo{year}{2018}), \bibinfo{pages}{1--26}.
\newblock


\bibitem[Kramer et~al\mbox{.}(2019)]%
        {kramer2019investigating}
\bibfield{author}{\bibinfo{person}{Jan-Niklas Kramer}, \bibinfo{person}{Florian K{\"u}nzler}, \bibinfo{person}{Varun Mishra}, \bibinfo{person}{Bastien Presset}, \bibinfo{person}{David Kotz}, \bibinfo{person}{Shawna Smith}, \bibinfo{person}{Urte Scholz}, \bibinfo{person}{Tobias Kowatsch}, {et~al\mbox{.}}} \bibinfo{year}{2019}\natexlab{}.
\newblock \showarticletitle{Investigating intervention components and exploring states of receptivity for a smartphone app to promote physical activity: protocol of a microrandomized trial}.
\newblock \bibinfo{journal}{\emph{JMIR research protocols}} \bibinfo{volume}{8}, \bibinfo{number}{1} (\bibinfo{year}{2019}), \bibinfo{pages}{e11540}.
\newblock


\bibitem[Kramer et~al\mbox{.}(2020a)]%
        {kramer2020components}
\bibfield{author}{\bibinfo{person}{Jan-Niklas Kramer}, \bibinfo{person}{Florian K{\"u}nzler}, \bibinfo{person}{Varun Mishra}, \bibinfo{person}{Shawna~N Smith}, \bibinfo{person}{David Kotz}, \bibinfo{person}{Urte Scholz}, \bibinfo{person}{Elgar Fleisch}, {and} \bibinfo{person}{Tobias Kowatsch}.} \bibinfo{year}{2020}\natexlab{a}.
\newblock \showarticletitle{Which components of a smartphone walking app help users to reach personalized step goals? Results from an optimization trial}.
\newblock \bibinfo{journal}{\emph{Annals of Behavioral Medicine}} \bibinfo{volume}{54}, \bibinfo{number}{7} (\bibinfo{year}{2020}), \bibinfo{pages}{518--528}.
\newblock


\bibitem[Kramer et~al\mbox{.}(2020b)]%
        {Krame_2020}
\bibfield{author}{\bibinfo{person}{Lean~L Kramer}, \bibinfo{person}{Silke ter Stal}, \bibinfo{person}{Bob~C Mulder}, \bibinfo{person}{Emely de Vet}, {and} \bibinfo{person}{Lex van Velsen}.} \bibinfo{year}{2020}\natexlab{b}.
\newblock \showarticletitle{Developing Embodied Conversational Agents for Coaching People in a Healthy Lifestyle: Scoping Review}.
\newblock \bibinfo{journal}{\emph{Journal of Medical Internet Research}} \bibinfo{volume}{22}, \bibinfo{number}{2} (\bibinfo{date}{Feb.} \bibinfo{year}{2020}), \bibinfo{pages}{e14058}.
\newblock
\showISSN{1439-4456}
\urldef\tempurl%
\url{https://doi.org/10.2196/14058}
\showDOI{\tempurl}


\bibitem[Kreuter et~al\mbox{.}(1999)]%
        {kreuter1999understanding}
\bibfield{author}{\bibinfo{person}{Matthew~W Kreuter}, \bibinfo{person}{Fiona~C Bull}, \bibinfo{person}{Eddie~M Clark}, {and} \bibinfo{person}{Debra~L Oswald}.} \bibinfo{year}{1999}\natexlab{}.
\newblock \showarticletitle{Understanding how people process health information: a comparison of tailored and nontailored weight-loss materials.}
\newblock \bibinfo{journal}{\emph{Health Psychology}} \bibinfo{volume}{18}, \bibinfo{number}{5} (\bibinfo{year}{1999}), \bibinfo{pages}{487}.
\newblock


\bibitem[Kross and Ayduk(2011)]%
        {kross2011making}
\bibfield{author}{\bibinfo{person}{Ethan Kross} {and} \bibinfo{person}{Ozlem Ayduk}.} \bibinfo{year}{2011}\natexlab{}.
\newblock \showarticletitle{Making meaning out of negative experiences by self-distancing}.
\newblock \bibinfo{journal}{\emph{Current directions in psychological science}} \bibinfo{volume}{20}, \bibinfo{number}{3} (\bibinfo{year}{2011}), \bibinfo{pages}{187--191}.
\newblock


\bibitem[Kumar and Epley(2018)]%
        {kumar2018undervaluing}
\bibfield{author}{\bibinfo{person}{Amit Kumar} {and} \bibinfo{person}{Nicholas Epley}.} \bibinfo{year}{2018}\natexlab{}.
\newblock \showarticletitle{Undervaluing gratitude: Expressers misunderstand the consequences of showing appreciation}.
\newblock \bibinfo{journal}{\emph{Psychological science}} \bibinfo{volume}{29}, \bibinfo{number}{9} (\bibinfo{year}{2018}), \bibinfo{pages}{1423--1435}.
\newblock


\bibitem[K{\"u}nzler et~al\mbox{.}(2019)]%
        {kunzler2019exploring}
\bibfield{author}{\bibinfo{person}{Florian K{\"u}nzler}, \bibinfo{person}{Varun Mishra}, \bibinfo{person}{Jan-Niklas Kramer}, \bibinfo{person}{David Kotz}, \bibinfo{person}{Elgar Fleisch}, {and} \bibinfo{person}{Tobias Kowatsch}.} \bibinfo{year}{2019}\natexlab{}.
\newblock \showarticletitle{Exploring the state-of-receptivity for mHealth interventions}.
\newblock \bibinfo{journal}{\emph{Proceedings of the ACM on Interactive, Mobile, Wearable and Ubiquitous Technologies}} \bibinfo{volume}{3}, \bibinfo{number}{4} (\bibinfo{year}{2019}), \bibinfo{pages}{1--27}.
\newblock


\bibitem[Kwak et~al\mbox{.}(2024)]%
        {kwak2024investigating}
\bibfield{author}{\bibinfo{person}{Daehyun Kwak}, \bibinfo{person}{Soobin Park}, \bibinfo{person}{Inha Cha}, \bibinfo{person}{Hankyung Kim}, {and} \bibinfo{person}{Youn-Kyung Lim}.} \bibinfo{year}{2024}\natexlab{}.
\newblock \showarticletitle{Investigating the Potential of Group Recommendation Systems As a Medium of Social Interactions: A Case of Spotify Blend Experiences between Two Users}. In \bibinfo{booktitle}{\emph{Proceedings of the CHI Conference on Human Factors in Computing Systems}}. \bibinfo{pages}{1--15}.
\newblock


\bibitem[Laban(2021)]%
        {laban2021perceptions}
\bibfield{author}{\bibinfo{person}{Guy Laban}.} \bibinfo{year}{2021}\natexlab{}.
\newblock \showarticletitle{Perceptions of anthropomorphism in a chatbot dialogue: the role of animacy and intelligence}. In \bibinfo{booktitle}{\emph{Proceedings of the 9th international conference on human-agent interaction}}. \bibinfo{pages}{305--310}.
\newblock


\bibitem[Laestadius et~al\mbox{.}(2022)]%
        {laestadius2022too}
\bibfield{author}{\bibinfo{person}{Linnea Laestadius}, \bibinfo{person}{Andrea Bishop}, \bibinfo{person}{Michael Gonzalez}, \bibinfo{person}{Diana Illen{\v{c}}{\'\i}k}, {and} \bibinfo{person}{Celeste Campos-Castillo}.} \bibinfo{year}{2022}\natexlab{}.
\newblock \showarticletitle{Too human and not human enough: A grounded theory analysis of mental health harms from emotional dependence on the social chatbot Replika}.
\newblock \bibinfo{journal}{\emph{new media \& society}} (\bibinfo{year}{2022}), \bibinfo{pages}{14614448221142007}.
\newblock


\bibitem[Laranjo et~al\mbox{.}(2018)]%
        {laranjo2018conversational}
\bibfield{author}{\bibinfo{person}{Liliana Laranjo}, \bibinfo{person}{Adam~G Dunn}, \bibinfo{person}{Huong~Ly Tong}, \bibinfo{person}{Ahmet~Baki Kocaballi}, \bibinfo{person}{Jessica Chen}, \bibinfo{person}{Rabia Bashir}, \bibinfo{person}{Didi Surian}, \bibinfo{person}{Blanca Gallego}, \bibinfo{person}{Farah Magrabi}, \bibinfo{person}{Annie~YS Lau}, {et~al\mbox{.}}} \bibinfo{year}{2018}\natexlab{}.
\newblock \showarticletitle{Conversational agents in healthcare: a systematic review}.
\newblock \bibinfo{journal}{\emph{Journal of the American Medical Informatics Association}} \bibinfo{volume}{25}, \bibinfo{number}{9} (\bibinfo{year}{2018}), \bibinfo{pages}{1248--1258}.
\newblock


\bibitem[Lee et~al\mbox{.}(2019)]%
        {Minha19}
\bibfield{author}{\bibinfo{person}{Minha Lee}, \bibinfo{person}{Sander Ackermans}, \bibinfo{person}{Nena van As}, \bibinfo{person}{Hanwen Chang}, \bibinfo{person}{Enzo Lucas}, {and} \bibinfo{person}{Wijnand IJsselsteijn}.} \bibinfo{year}{2019}\natexlab{}.
\newblock \showarticletitle{Caring for Vincent: A Chatbot for Self-Compassion}. In \bibinfo{booktitle}{\emph{Proceedings of the 2019 CHI Conference on Human Factors in Computing Systems}} (Glasgow, Scotland Uk) \emph{(\bibinfo{series}{CHI '19})}. \bibinfo{publisher}{Association for Computing Machinery}, \bibinfo{address}{New York, NY, USA}, \bibinfo{pages}{1–13}.
\newblock
\showISBNx{9781450359702}
\urldef\tempurl%
\url{https://doi.org/10.1145/3290605.3300932}
\showDOI{\tempurl}


\bibitem[Lee et~al\mbox{.}(2020a)]%
        {Lee20}
\bibfield{author}{\bibinfo{person}{Yi-Chieh Lee}, \bibinfo{person}{Naomi Yamashita}, {and} \bibinfo{person}{Yun Huang}.} \bibinfo{year}{2020}\natexlab{a}.
\newblock \showarticletitle{Designing a Chatbot as a Mediator for Promoting Deep Self-Disclosure to a Real Mental Health Professional}.
\newblock \bibinfo{journal}{\emph{Proc. ACM Hum.-Comput. Interact.}} \bibinfo{volume}{4}, \bibinfo{number}{CSCW1}, Article \bibinfo{articleno}{31} (\bibinfo{date}{may} \bibinfo{year}{2020}), \bibinfo{numpages}{27}~pages.
\newblock
\urldef\tempurl%
\url{https://doi.org/10.1145/3392836}
\showDOI{\tempurl}


\bibitem[Lee et~al\mbox{.}(2020b)]%
        {Lee20chi}
\bibfield{author}{\bibinfo{person}{Yi-Chieh Lee}, \bibinfo{person}{Naomi Yamashita}, \bibinfo{person}{Yun Huang}, {and} \bibinfo{person}{Wai Fu}.} \bibinfo{year}{2020}\natexlab{b}.
\newblock \showarticletitle{"I Hear You, I Feel You": Encouraging Deep Self-disclosure through a Chatbot}. In \bibinfo{booktitle}{\emph{Proceedings of the 2020 CHI Conference on Human Factors in Computing Systems}} (<conf-loc>, <city>Honolulu</city>, <state>HI</state>, <country>USA</country>, </conf-loc>) \emph{(\bibinfo{series}{CHI '20})}. \bibinfo{publisher}{Association for Computing Machinery}, \bibinfo{address}{New York, NY, USA}, \bibinfo{pages}{1–12}.
\newblock
\showISBNx{9781450367080}
\urldef\tempurl%
\url{https://doi.org/10.1145/3313831.3376175}
\showDOI{\tempurl}


\bibitem[Lee et~al\mbox{.}(2023)]%
        {lee2023chain}
\bibfield{author}{\bibinfo{person}{Yoon~Kyung Lee}, \bibinfo{person}{Inju Lee}, \bibinfo{person}{Minjung Shin}, \bibinfo{person}{Seoyeon Bae}, {and} \bibinfo{person}{Sowon Hahn}.} \bibinfo{year}{2023}\natexlab{}.
\newblock \bibinfo{title}{Chain of Empathy: Enhancing Empathetic Response of Large Language Models Based on Psychotherapy Models}.
\newblock
\newblock
\showeprint[arxiv]{2311.04915}~[cs.CL]


\bibitem[Li et~al\mbox{.}(2023b)]%
        {li2023systematic}
\bibfield{author}{\bibinfo{person}{Han Li}, \bibinfo{person}{Renwen Zhang}, \bibinfo{person}{Yi-Chieh Lee}, \bibinfo{person}{Robert~E Kraut}, {and} \bibinfo{person}{David~C Mohr}.} \bibinfo{year}{2023}\natexlab{b}.
\newblock \showarticletitle{Systematic review and meta-analysis of AI-based conversational agents for promoting mental health and well-being}.
\newblock \bibinfo{journal}{\emph{NPJ Digital Medicine}} \bibinfo{volume}{6}, \bibinfo{number}{1} (\bibinfo{year}{2023}), \bibinfo{pages}{236}.
\newblock


\bibitem[Li et~al\mbox{.}(2023a)]%
        {li2023influence}
\bibfield{author}{\bibinfo{person}{Qingchuan Li}, \bibinfo{person}{Yan Luximon}, {and} \bibinfo{person}{Jiaxin Zhang}.} \bibinfo{year}{2023}\natexlab{a}.
\newblock \showarticletitle{The Influence of Anthropomorphic Cues on Patients’ Perceived Anthropomorphism, Social Presence, Trust Building, and Acceptance of Health Care Conversational Agents: Within-Subject Web-Based Experiment}.
\newblock \bibinfo{journal}{\emph{Journal of Medical Internet Research}}  \bibinfo{volume}{25} (\bibinfo{year}{2023}), \bibinfo{pages}{e44479}.
\newblock


\bibitem[Liu et~al\mbox{.}(2022)]%
        {liu2022using}
\bibfield{author}{\bibinfo{person}{Hao Liu}, \bibinfo{person}{Huaming Peng}, \bibinfo{person}{Xingyu Song}, \bibinfo{person}{Chenzi Xu}, {and} \bibinfo{person}{Meng Zhang}.} \bibinfo{year}{2022}\natexlab{}.
\newblock \showarticletitle{Using AI chatbots to provide self-help depression interventions for university students: A randomized trial of effectiveness}.
\newblock \bibinfo{journal}{\emph{Internet Interventions}}  \bibinfo{volume}{27} (\bibinfo{year}{2022}), \bibinfo{pages}{100495}.
\newblock


\bibitem[Liu et~al\mbox{.}(2023b)]%
        {liu2023chatcounselor}
\bibfield{author}{\bibinfo{person}{June~M. Liu}, \bibinfo{person}{Donghao Li}, \bibinfo{person}{He Cao}, \bibinfo{person}{Tianhe Ren}, \bibinfo{person}{Zeyi Liao}, {and} \bibinfo{person}{Jiamin Wu}.} \bibinfo{year}{2023}\natexlab{b}.
\newblock \bibinfo{title}{ChatCounselor: A Large Language Models for Mental Health Support}.
\newblock
\newblock
\showeprint[arxiv]{2309.15461}~[cs.CL]


\bibitem[Liu et~al\mbox{.}(2023a)]%
        {liu2023taskadaptive}
\bibfield{author}{\bibinfo{person}{Siyang Liu}, \bibinfo{person}{Naihao Deng}, \bibinfo{person}{Sahand Sabour}, \bibinfo{person}{Yilin Jia}, \bibinfo{person}{Minlie Huang}, {and} \bibinfo{person}{Rada Mihalcea}.} \bibinfo{year}{2023}\natexlab{a}.
\newblock \bibinfo{title}{Task-Adaptive Tokenization: Enhancing Long-Form Text Generation Efficacy in Mental Health and Beyond}.
\newblock
\newblock
\showeprint[arxiv]{2310.05317}~[cs.CL]


\bibitem[Loh and Raamkumar(2023)]%
        {loh2023harnessing}
\bibfield{author}{\bibinfo{person}{Siyuan~Brandon Loh} {and} \bibinfo{person}{Aravind~Sesagiri Raamkumar}.} \bibinfo{year}{2023}\natexlab{}.
\newblock \bibinfo{title}{Harnessing Large Language Models' Empathetic Response Generation Capabilities for Online Mental Health Counselling Support}.
\newblock
\newblock
\showeprint[arxiv]{2310.08017}~[cs.CL]


\bibitem[Luger and Sellen(2016)]%
        {luger2016like}
\bibfield{author}{\bibinfo{person}{Ewa Luger} {and} \bibinfo{person}{Abigail Sellen}.} \bibinfo{year}{2016}\natexlab{}.
\newblock \showarticletitle{Like having a really bad PA: The gulf between user expectation and experience of conversational agents}. In \bibinfo{booktitle}{\emph{Proceedings of the 2016 CHI Conference on Human Factors in Computing Systems}}. ACM, \bibinfo{pages}{5286--5297}.
\newblock


\bibitem[Ly et~al\mbox{.}(2017)]%
        {Ly_Ly_Andersson_2017}
\bibfield{author}{\bibinfo{person}{Kien~Hoa Ly}, \bibinfo{person}{Ann-Marie Ly}, {and} \bibinfo{person}{Gerhard Andersson}.} \bibinfo{year}{2017}\natexlab{}.
\newblock \showarticletitle{A fully automated conversational agent for promoting mental well-being: A pilot RCT using mixed methods}.
\newblock   \bibinfo{volume}{10} (\bibinfo{date}{Dec.} \bibinfo{year}{2017}), \bibinfo{pages}{39–46}.
\newblock
\showISSN{2214-7829}
\urldef\tempurl%
\url{https://doi.org/10.1016/j.invent.2017.10.002}
\showDOI{\tempurl}


\bibitem[Lyubomirsky et~al\mbox{.}(2011)]%
        {lyubomirsky2011becoming}
\bibfield{author}{\bibinfo{person}{Sonja Lyubomirsky}, \bibinfo{person}{Rene Dickerhoof}, \bibinfo{person}{Julia~K Boehm}, {and} \bibinfo{person}{Kennon~M Sheldon}.} \bibinfo{year}{2011}\natexlab{}.
\newblock \showarticletitle{Becoming happier takes both a will and a proper way: an experimental longitudinal intervention to boost well-being.}
\newblock \bibinfo{journal}{\emph{Emotion}} \bibinfo{volume}{11}, \bibinfo{number}{2} (\bibinfo{year}{2011}), \bibinfo{pages}{391}.
\newblock


\bibitem[Lyubomirsky and Layous(2013)]%
        {lyubomirsky2013simple}
\bibfield{author}{\bibinfo{person}{Sonja Lyubomirsky} {and} \bibinfo{person}{Kristin Layous}.} \bibinfo{year}{2013}\natexlab{}.
\newblock \showarticletitle{How do simple positive activities increase well-being?}
\newblock \bibinfo{journal}{\emph{Current directions in psychological science}} \bibinfo{volume}{22}, \bibinfo{number}{1} (\bibinfo{year}{2013}), \bibinfo{pages}{57--62}.
\newblock


\bibitem[Ma et~al\mbox{.}(2023)]%
        {ma2023understanding}
\bibfield{author}{\bibinfo{person}{Zilin Ma}, \bibinfo{person}{Yiyang Mei}, {and} \bibinfo{person}{Zhaoyuan Su}.} \bibinfo{year}{2023}\natexlab{}.
\newblock \bibinfo{title}{Understanding the Benefits and Challenges of Using Large Language Model-based Conversational Agents for Mental Well-being Support}.
\newblock
\newblock
\showeprint[arxiv]{2307.15810}~[cs.HC]


\bibitem[Maher et~al\mbox{.}(2020)]%
        {maher2020physical}
\bibfield{author}{\bibinfo{person}{Carol~Ann Maher}, \bibinfo{person}{Courtney~Rose Davis}, \bibinfo{person}{Rachel~Grace Curtis}, \bibinfo{person}{Camille~Elizabeth Short}, {and} \bibinfo{person}{Karen~Joy Murphy}.} \bibinfo{year}{2020}\natexlab{}.
\newblock \showarticletitle{A physical activity and diet program delivered by artificially intelligent virtual health coach: proof-of-concept study}.
\newblock \bibinfo{journal}{\emph{JMIR mHealth and uHealth}} \bibinfo{volume}{8}, \bibinfo{number}{7} (\bibinfo{year}{2020}), \bibinfo{pages}{e17558}.
\newblock


\bibitem[Mahmood et~al\mbox{.}(2023)]%
        {mahmood2023llm}
\bibfield{author}{\bibinfo{person}{Amama Mahmood}, \bibinfo{person}{Junxiang Wang}, \bibinfo{person}{Bingsheng Yao}, \bibinfo{person}{Dakuo Wang}, {and} \bibinfo{person}{Chien-Ming Huang}.} \bibinfo{year}{2023}\natexlab{}.
\newblock \showarticletitle{LLM-Powered Conversational Voice Assistants: Interaction Patterns, Opportunities, Challenges, and Design Guidelines}.
\newblock \bibinfo{journal}{\emph{arXiv preprint arXiv:2309.13879}} (\bibinfo{year}{2023}).
\newblock


\bibitem[McKee et~al\mbox{.}(2023)]%
        {mckee2023warmth}
\bibfield{author}{\bibinfo{person}{Kathryn~R McKee}, \bibinfo{person}{Xi Bai}, {and} \bibinfo{person}{Susan~T Fiske}.} \bibinfo{year}{2023}\natexlab{}.
\newblock \showarticletitle{Humans perceive warmth and competence in artificial intelligence}.
\newblock \bibinfo{journal}{\emph{iScience}} \bibinfo{volume}{26}, \bibinfo{number}{8} (\bibinfo{year}{2023}), \bibinfo{pages}{1--16}.
\newblock


\bibitem[McTear(2018)]%
        {mctear2018conversational}
\bibfield{author}{\bibinfo{person}{Michael McTear}.} \bibinfo{year}{2018}\natexlab{}.
\newblock \showarticletitle{Conversational modelling for chatbots: current approaches and future directions}.
\newblock \bibinfo{journal}{\emph{Studientexte zur Sprachkommunikation: Elektronische Sprachsignalverarbeitung}} (\bibinfo{year}{2018}), \bibinfo{pages}{175--185}.
\newblock


\bibitem[Michalco et~al\mbox{.}(2015)]%
        {michalco2015relation}
\bibfield{author}{\bibinfo{person}{Jan Michalco}, \bibinfo{person}{Jakob~Grue Simonsen}, {and} \bibinfo{person}{Kasper Hornb{\ae}k}.} \bibinfo{year}{2015}\natexlab{}.
\newblock \showarticletitle{An exploration of the relation between expectations and user experience}.
\newblock \bibinfo{journal}{\emph{International Journal of Human-Computer Interaction}} \bibinfo{volume}{31}, \bibinfo{number}{9} (\bibinfo{year}{2015}), \bibinfo{pages}{603--617}.
\newblock


\bibitem[Mori(1970)]%
        {mori1970uncanny}
\bibfield{author}{\bibinfo{person}{Masahiro Mori}.} \bibinfo{year}{1970}\natexlab{}.
\newblock \showarticletitle{The uncanny valley: the original essay by Masahiro Mori}.
\newblock \bibinfo{journal}{\emph{Ieee Spectrum}}  \bibinfo{volume}{6} (\bibinfo{year}{1970}), \bibinfo{pages}{1--6}.
\newblock


\bibitem[Nass and Moon(2000)]%
        {nass2000machines}
\bibfield{author}{\bibinfo{person}{Clifford Nass} {and} \bibinfo{person}{Youngme Moon}.} \bibinfo{year}{2000}\natexlab{}.
\newblock \showarticletitle{Machines and mindlessness: Social responses to computers}.
\newblock \bibinfo{journal}{\emph{Journal of social issues}} \bibinfo{volume}{56}, \bibinfo{number}{1} (\bibinfo{year}{2000}), \bibinfo{pages}{81--103}.
\newblock


\bibitem[Nass et~al\mbox{.}(1994)]%
        {nass1994computers}
\bibfield{author}{\bibinfo{person}{Clifford Nass}, \bibinfo{person}{Jonathan Steuer}, {and} \bibinfo{person}{Ellen~R Tauber}.} \bibinfo{year}{1994}\natexlab{}.
\newblock \showarticletitle{Computers are social actors}. In \bibinfo{booktitle}{\emph{Proceedings of the SIGCHI conference on Human factors in computing systems}}. \bibinfo{pages}{72--78}.
\newblock


\bibitem[Oh et~al\mbox{.}(2021)]%
        {Oh2021ASR}
\bibfield{author}{\bibinfo{person}{Yoo~Jung Oh}, \bibinfo{person}{Jingwen Zhang}, \bibinfo{person}{Min Fang}, {and} \bibinfo{person}{Yoshimi Fukuoka}.} \bibinfo{year}{2021}\natexlab{}.
\newblock \showarticletitle{A systematic review of artificial intelligence chatbots for promoting physical activity, healthy diet, and weight loss}.
\newblock \bibinfo{journal}{\emph{International Journal of Behavioral Nutrition and Physical Activity}}  \bibinfo{volume}{18} (\bibinfo{year}{2021}), \bibinfo{pages}{1--25}.
\newblock
\urldef\tempurl%
\url{https://api.semanticscholar.org/CorpusID:245014899}
\showURL{%
\tempurl}


\bibitem[Organization(2023)]%
        {who2023}
\bibfield{author}{\bibinfo{person}{World~Health Organization}.} \bibinfo{year}{2023}\natexlab{}.
\newblock \bibinfo{title}{Mental health and COVID-19: Early evidence of the pandemic’s impact}.
\newblock
\newblock
\urldef\tempurl%
\url{https://www.who.int/publications/i/item/WHO-2019-nCoV-Sci_Brief-Mental_health-2022.1}
\showURL{%
\tempurl}
\newblock
\shownote{Accessed: 2023-09-10}.


\bibitem[Park et~al\mbox{.}(2012)]%
        {park2012law}
\bibfield{author}{\bibinfo{person}{Eunil Park}, \bibinfo{person}{Dallae Jin}, {and} \bibinfo{person}{Angel~P Del~Pobil}.} \bibinfo{year}{2012}\natexlab{}.
\newblock \showarticletitle{The law of attraction in human-robot interaction}.
\newblock \bibinfo{journal}{\emph{International Journal of Advanced Robotic Systems}} \bibinfo{volume}{9}, \bibinfo{number}{2} (\bibinfo{year}{2012}), \bibinfo{pages}{35}.
\newblock


\bibitem[Penedo and Dahn(2005)]%
        {Penedo2005ExerciseAW}
\bibfield{author}{\bibinfo{person}{Frank~J. Penedo} {and} \bibinfo{person}{Jason~R. Dahn}.} \bibinfo{year}{2005}\natexlab{}.
\newblock \showarticletitle{Exercise and well-being: a review of mental and physical health benefits associated with physical activity}.
\newblock \bibinfo{journal}{\emph{Current Opinion in Psychiatry}}  \bibinfo{volume}{18} (\bibinfo{year}{2005}), \bibinfo{pages}{189--193}.
\newblock
\urldef\tempurl%
\url{https://doi.org/10.1097/00001504-200503000-00013}
\showDOI{\tempurl}


\bibitem[Pennebaker and Chung(2011)]%
        {pennebaker2011expressive}
\bibfield{author}{\bibinfo{person}{James~W Pennebaker} {and} \bibinfo{person}{Cindy~K Chung}.} \bibinfo{year}{2011}\natexlab{}.
\newblock \showarticletitle{Expressive writing: Connections to physical and mental health.(2011)}.
\newblock  (\bibinfo{year}{2011}).
\newblock


\bibitem[Perski et~al\mbox{.}(2017)]%
        {perski2017conceptualising}
\bibfield{author}{\bibinfo{person}{Olga Perski}, \bibinfo{person}{Ann Blandford}, \bibinfo{person}{Robert West}, {and} \bibinfo{person}{Susan Michie}.} \bibinfo{year}{2017}\natexlab{}.
\newblock \showarticletitle{Conceptualising engagement with digital behaviour change interventions: a systematic review using principles from critical interpretive synthesis}.
\newblock \bibinfo{journal}{\emph{Translational behavioral medicine}} \bibinfo{volume}{7}, \bibinfo{number}{2} (\bibinfo{year}{2017}), \bibinfo{pages}{254--267}.
\newblock


\bibitem[Perski et~al\mbox{.}(2019)]%
        {Perski2019DoesTA}
\bibfield{author}{\bibinfo{person}{Olga Perski}, \bibinfo{person}{David Crane}, \bibinfo{person}{Emma Beard}, {and} \bibinfo{person}{Jamie Brown}.} \bibinfo{year}{2019}\natexlab{}.
\newblock \showarticletitle{Does the addition of a supportive chatbot promote user engagement with a smoking cessation app? An experimental study}.
\newblock \bibinfo{journal}{\emph{Digital Health}}  \bibinfo{volume}{5} (\bibinfo{year}{2019}).
\newblock
\urldef\tempurl%
\url{https://api.semanticscholar.org/CorpusID:204737594}
\showURL{%
\tempurl}


\bibitem[Piao et~al\mbox{.}(2020)]%
        {piao2020use}
\bibfield{author}{\bibinfo{person}{Meihua Piao}, \bibinfo{person}{Hyeongju Ryu}, \bibinfo{person}{Hyeongsuk Lee}, \bibinfo{person}{Jeongeun Kim}, {et~al\mbox{.}}} \bibinfo{year}{2020}\natexlab{}.
\newblock \showarticletitle{Use of the healthy lifestyle coaching chatbot app to promote stair-climbing habits among office workers: exploratory randomized controlled trial}.
\newblock \bibinfo{journal}{\emph{JMIR mHealth and uHealth}} \bibinfo{volume}{8}, \bibinfo{number}{5} (\bibinfo{year}{2020}), \bibinfo{pages}{e15085}.
\newblock


\bibitem[Porcheron et~al\mbox{.}(2018)]%
        {porcheron2018voice}
\bibfield{author}{\bibinfo{person}{Martin Porcheron}, \bibinfo{person}{Joel~E Fischer}, \bibinfo{person}{Stuart Reeves}, {and} \bibinfo{person}{Sarah Sharples}.} \bibinfo{year}{2018}\natexlab{}.
\newblock \showarticletitle{Voice interfaces in everyday life}. In \bibinfo{booktitle}{\emph{proceedings of the 2018 CHI conference on human factors in computing systems}}. \bibinfo{pages}{1--12}.
\newblock


\bibitem[Pradhan and Lazar(2021)]%
        {Pradhan2021}
\bibfield{author}{\bibinfo{person}{Alisha Pradhan} {and} \bibinfo{person}{Amanda Lazar}.} \bibinfo{year}{2021}\natexlab{}.
\newblock \showarticletitle{Hey Google, Do You Have a Personality? Designing Personality and Personas for Conversational Agents}. In \bibinfo{booktitle}{\emph{Proceedings of the 3rd Conference on Conversational User Interfaces}} (Bilbao (online), Spain) \emph{(\bibinfo{series}{CUI '21})}. \bibinfo{publisher}{Association for Computing Machinery}, \bibinfo{address}{New York, NY, USA}, Article \bibinfo{articleno}{12}, \bibinfo{numpages}{4}~pages.
\newblock
\showISBNx{9781450389983}
\urldef\tempurl%
\url{https://doi.org/10.1145/3469595.3469607}
\showDOI{\tempurl}


\bibitem[Prochaska and Velicer(1997)]%
        {Prochaska1997}
\bibfield{author}{\bibinfo{person}{James~O. Prochaska} {and} \bibinfo{person}{Wayne~F. Velicer}.} \bibinfo{year}{1997}\natexlab{}.
\newblock \showarticletitle{The transtheoretical model of health behavior change}.
\newblock \bibinfo{journal}{\emph{American Journal of Health Promotion}} \bibinfo{volume}{12}, \bibinfo{number}{1} (\bibinfo{date}{Sep-Oct} \bibinfo{year}{1997}), \bibinfo{pages}{38--48}.
\newblock
\urldef\tempurl%
\url{https://doi.org/10.4278/0890-1171-12.1.38}
\showDOI{\tempurl}


\bibitem[Reeves et~al\mbox{.}(2018)]%
        {reeves2018not}
\bibfield{author}{\bibinfo{person}{Stuart Reeves}, \bibinfo{person}{Martin Porcheron}, {and} \bibinfo{person}{Joel Fischer}.} \bibinfo{year}{2018}\natexlab{}.
\newblock \showarticletitle{'This is not what we wanted' designing for conversation with voice interfaces}.
\newblock \bibinfo{journal}{\emph{Interactions}} \bibinfo{volume}{26}, \bibinfo{number}{1} (\bibinfo{year}{2018}), \bibinfo{pages}{46--51}.
\newblock


\bibitem[Rohani et~al\mbox{.}(2020)]%
        {Rohani2020MUBSAP}
\bibfield{author}{\bibinfo{person}{Darius~Adam Rohani}, \bibinfo{person}{Andrea~Quemada Lopategui}, \bibinfo{person}{Nanna Tuxen}, \bibinfo{person}{Maria Faurholt-Jepsen}, \bibinfo{person}{Lars~Vedel Kessing}, {and} \bibinfo{person}{Jakob~Eyvind Bardram}.} \bibinfo{year}{2020}\natexlab{}.
\newblock \showarticletitle{MUBS: A Personalized Recommender System for Behavioral Activation in Mental Health}.
\newblock \bibinfo{journal}{\emph{Proceedings of the 2020 CHI Conference on Human Factors in Computing Systems}} (\bibinfo{year}{2020}).
\newblock
\urldef\tempurl%
\url{https://api.semanticscholar.org/CorpusID:218482533}
\showURL{%
\tempurl}


\bibitem[Ryan and Deci(2000)]%
        {Ryan2000}
\bibfield{author}{\bibinfo{person}{Richard~M. Ryan} {and} \bibinfo{person}{Edward~L. Deci}.} \bibinfo{year}{2000}\natexlab{}.
\newblock \showarticletitle{Self-determination theory and the facilitation of intrinsic motivation, social development, and well-being}.
\newblock \bibinfo{journal}{\emph{American Psychologist}} \bibinfo{volume}{55}, \bibinfo{number}{1} (\bibinfo{year}{2000}), \bibinfo{pages}{68--78}.
\newblock
\urldef\tempurl%
\url{https://doi.org/10.1037/0003-066X.55.1.68}
\showDOI{\tempurl}


\bibitem[Sanchez et~al\mbox{.}(2014)]%
        {sanchez2014acceptability}
\bibfield{author}{\bibinfo{person}{Rebecca~Polley Sanchez}, \bibinfo{person}{Chelsea~M Bartel}, \bibinfo{person}{Emily Brown}, {and} \bibinfo{person}{Melissa DeRosier}.} \bibinfo{year}{2014}\natexlab{}.
\newblock \showarticletitle{The acceptability and efficacy of an intelligent social tutoring system}.
\newblock \bibinfo{journal}{\emph{Computers \& Education}}  \bibinfo{volume}{78} (\bibinfo{year}{2014}), \bibinfo{pages}{321--332}.
\newblock


\bibitem[Schueller et~al\mbox{.}(2013)]%
        {Schueller2013}
\bibfield{author}{\bibinfo{person}{Stephen~M. Schueller}, \bibinfo{person}{Ricardo~F. Mu{\~n}oz}, {and} \bibinfo{person}{David~C. Mohr}.} \bibinfo{year}{2013}\natexlab{}.
\newblock \showarticletitle{Realizing the Potential of Behavioral Intervention Technologies}.
\newblock \bibinfo{journal}{\emph{Current Directions in Psychological Science}} \bibinfo{volume}{22}, \bibinfo{number}{6} (\bibinfo{year}{2013}), \bibinfo{pages}{478--483}.
\newblock
\urldef\tempurl%
\url{https://doi.org/10.1177/0963721413495872}
\showDOI{\tempurl}


\bibitem[Seeger et~al\mbox{.}(2018)]%
        {seeger2018designing}
\bibfield{author}{\bibinfo{person}{Anna-Maria Seeger}, \bibinfo{person}{Jella Pfeiffer}, {and} \bibinfo{person}{Armin Heinzl}.} \bibinfo{year}{2018}\natexlab{}.
\newblock \showarticletitle{Designing anthropomorphic conversational agents: Development and empirical evaluation of a design framework}.
\newblock  (\bibinfo{year}{2018}).
\newblock


\bibitem[Seligman et~al\mbox{.}(2005)]%
        {seligman2005positive}
\bibfield{author}{\bibinfo{person}{Martin~EP Seligman}, \bibinfo{person}{Tracy~A Steen}, \bibinfo{person}{Nansook Park}, {and} \bibinfo{person}{Christopher Peterson}.} \bibinfo{year}{2005}\natexlab{}.
\newblock \showarticletitle{Positive psychology progress: empirical validation of interventions.}
\newblock \bibinfo{journal}{\emph{American psychologist}} \bibinfo{volume}{60}, \bibinfo{number}{5} (\bibinfo{year}{2005}), \bibinfo{pages}{410}.
\newblock


\bibitem[Seligman(2011)]%
        {seligman2011}
\bibfield{author}{\bibinfo{person}{Martin E.~P. Seligman}.} \bibinfo{year}{2011}\natexlab{}.
\newblock \bibinfo{booktitle}{\emph{Flourish: A visionary new understanding of happiness and well-being}}.
\newblock \bibinfo{publisher}{Free Press}.
\newblock


\bibitem[Sheldon and Lyubomirsky(2006)]%
        {sheldon2006increase}
\bibfield{author}{\bibinfo{person}{Kennon~M Sheldon} {and} \bibinfo{person}{Sonja Lyubomirsky}.} \bibinfo{year}{2006}\natexlab{}.
\newblock \showarticletitle{How to increase and sustain positive emotion: The effects of expressing gratitude and visualizing best possible selves}.
\newblock \bibinfo{journal}{\emph{The journal of positive psychology}} \bibinfo{volume}{1}, \bibinfo{number}{2} (\bibinfo{year}{2006}), \bibinfo{pages}{73--82}.
\newblock


\bibitem[Singh et~al\mbox{.}(2023)]%
        {Singh2023SystematicRA}
\bibfield{author}{\bibinfo{person}{Ben Singh}, \bibinfo{person}{Timothy Olds}, \bibinfo{person}{Jacinta Brinsley}, \bibinfo{person}{Dorothea Dumuid}, \bibinfo{person}{Rosa Virgara}, \bibinfo{person}{Lisa Matricciani}, \bibinfo{person}{Amanda Watson}, \bibinfo{person}{Kim Szeto}, \bibinfo{person}{Emily Eglitis}, \bibinfo{person}{Aaron Miatke}, \bibinfo{person}{Catherine E~M Simpson}, \bibinfo{person}{Corneel Vandelanotte}, {and} \bibinfo{person}{Carol~Ann Maher}.} \bibinfo{year}{2023}\natexlab{}.
\newblock \showarticletitle{Systematic review and meta-analysis of the effectiveness of chatbots on lifestyle behaviours}.
\newblock \bibinfo{journal}{\emph{NPJ Digital Medicine}}  \bibinfo{volume}{6} (\bibinfo{year}{2023}).
\newblock
\urldef\tempurl%
\url{https://api.semanticscholar.org/CorpusID:259240488}
\showURL{%
\tempurl}


\bibitem[Song et~al\mbox{.}(2024)]%
        {Song2024TheTC}
\bibfield{author}{\bibinfo{person}{Inhwa Song}, \bibinfo{person}{Sachin~R. Pendse}, \bibinfo{person}{Neha Kumar}, {and} \bibinfo{person}{Munmun~De Choudhury}.} \bibinfo{year}{2024}\natexlab{}.
\newblock \showarticletitle{The Typing Cure: Experiences with Large Language Model Chatbots for Mental Health Support}.
\newblock \bibinfo{journal}{\emph{ArXiv}}  \bibinfo{volume}{abs/2401.14362} (\bibinfo{year}{2024}).
\newblock
\urldef\tempurl%
\url{https://api.semanticscholar.org/CorpusID:267211682}
\showURL{%
\tempurl}


\bibitem[Stephens et~al\mbox{.}(2019)]%
        {stephens2019feasibility}
\bibfield{author}{\bibinfo{person}{Taylor~N Stephens}, \bibinfo{person}{Angela Joerin}, \bibinfo{person}{Michiel Rauws}, {and} \bibinfo{person}{Lloyd~N Werk}.} \bibinfo{year}{2019}\natexlab{}.
\newblock \showarticletitle{Feasibility of pediatric obesity and prediabetes treatment support through Tess, the AI behavioral coaching chatbot}.
\newblock \bibinfo{journal}{\emph{Translational behavioral medicine}} \bibinfo{volume}{9}, \bibinfo{number}{3} (\bibinfo{year}{2019}), \bibinfo{pages}{440--447}.
\newblock


\bibitem[Stever(2017)]%
        {stever2017parasocial}
\bibfield{author}{\bibinfo{person}{Gayle~S Stever}.} \bibinfo{year}{2017}\natexlab{}.
\newblock \showarticletitle{Parasocial theory: Concepts and measures}.
\newblock \bibinfo{journal}{\emph{The international encyclopedia of media effects}} (\bibinfo{year}{2017}), \bibinfo{pages}{1--12}.
\newblock


\bibitem[Strohmann et~al\mbox{.}(2023)]%
        {strohmann2023toward}
\bibfield{author}{\bibinfo{person}{Timo Strohmann}, \bibinfo{person}{Dominik Siemon}, \bibinfo{person}{Bijan Khosrawi-Rad}, {and} \bibinfo{person}{Susanne Robra-Bissantz}.} \bibinfo{year}{2023}\natexlab{}.
\newblock \showarticletitle{Toward a design theory for virtual companionship}.
\newblock \bibinfo{journal}{\emph{Human--Computer Interaction}} \bibinfo{volume}{38}, \bibinfo{number}{3-4} (\bibinfo{year}{2023}), \bibinfo{pages}{194--234}.
\newblock


\bibitem[Torous and Blease(2024)]%
        {torous2024generative}
\bibfield{author}{\bibinfo{person}{John Torous} {and} \bibinfo{person}{Charlotte Blease}.} \bibinfo{year}{2024}\natexlab{}.
\newblock \showarticletitle{Generative artificial intelligence in mental health care: potential benefits and current challenges}.
\newblock \bibinfo{journal}{\emph{World Psychiatry}} \bibinfo{volume}{23}, \bibinfo{number}{1} (\bibinfo{year}{2024}), \bibinfo{pages}{1}.
\newblock


\bibitem[Tosti et~al\mbox{.}(2024)]%
        {tosti2024using}
\bibfield{author}{\bibinfo{person}{Beatrice Tosti}, \bibinfo{person}{Stefano Corrado}, {and} \bibinfo{person}{Stefania Mancone}.} \bibinfo{year}{2024}\natexlab{}.
\newblock \showarticletitle{Using Chatbots and Conversational Agents for the Promotion of Well-being and Mental Health in Adolescents: Limitations and Perspectives.}
\newblock \bibinfo{journal}{\emph{Journal of Inclusive Methodology and Technology in Learning and Teaching}} \bibinfo{volume}{4}, \bibinfo{number}{1} (\bibinfo{year}{2024}).
\newblock


\bibitem[Tsai et~al\mbox{.}(2006)]%
        {tsai2006cultural}
\bibfield{author}{\bibinfo{person}{Jeanne~L. Tsai}, \bibinfo{person}{Brian Knutson}, {and} \bibinfo{person}{Helene~H. Fung}.} \bibinfo{year}{2006}\natexlab{}.
\newblock \showarticletitle{Cultural variation in affect valuation}.
\newblock \bibinfo{journal}{\emph{Journal of Personality and Social Psychology}} \bibinfo{volume}{90}, \bibinfo{number}{2} (\bibinfo{year}{2006}), \bibinfo{pages}{288--307}.
\newblock


\bibitem[Turan et~al\mbox{.}(2016)]%
        {turan2016older}
\bibfield{author}{\bibinfo{person}{Bulent Turan}, \bibinfo{person}{Tamara Sims}, \bibinfo{person}{Sasha~E Best}, {and} \bibinfo{person}{Laura~L Carstensen}.} \bibinfo{year}{2016}\natexlab{}.
\newblock \showarticletitle{Older age may offset genetic influence on affect: The COMT polymorphism and affective well-being across the life span.}
\newblock \bibinfo{journal}{\emph{Psychology and Aging}} \bibinfo{volume}{31}, \bibinfo{number}{3} (\bibinfo{year}{2016}), \bibinfo{pages}{287}.
\newblock


\bibitem[Vaidyam et~al\mbox{.}(2019)]%
        {vaidyam2019chatbots}
\bibfield{author}{\bibinfo{person}{Aditya~Nrusimha Vaidyam}, \bibinfo{person}{Hannah Wisniewski}, \bibinfo{person}{John~David Halamka}, \bibinfo{person}{Matcheri~S Kashavan}, {and} \bibinfo{person}{John~Blake Torous}.} \bibinfo{year}{2019}\natexlab{}.
\newblock \showarticletitle{Chatbots and conversational agents in mental health: a review of the psychiatric landscape}.
\newblock \bibinfo{journal}{\emph{The Canadian Journal of Psychiatry}} \bibinfo{volume}{64}, \bibinfo{number}{7} (\bibinfo{year}{2019}), \bibinfo{pages}{456--464}.
\newblock


\bibitem[van Heerden et~al\mbox{.}(2023)]%
        {van2023global}
\bibfield{author}{\bibinfo{person}{Alastair~C van Heerden}, \bibinfo{person}{Julia~R Pozuelo}, {and} \bibinfo{person}{Brandon~A Kohrt}.} \bibinfo{year}{2023}\natexlab{}.
\newblock \showarticletitle{Global mental health services and the impact of artificial intelligence--powered large language models}.
\newblock \bibinfo{journal}{\emph{JAMA psychiatry}} \bibinfo{volume}{80}, \bibinfo{number}{7} (\bibinfo{year}{2023}), \bibinfo{pages}{662--664}.
\newblock


\bibitem[VanderWeele(2017)]%
        {VanderWeele2017}
\bibfield{author}{\bibinfo{person}{Tyler~J. VanderWeele}.} \bibinfo{year}{2017}\natexlab{}.
\newblock \showarticletitle{On the promotion of human flourishing}.
\newblock \bibinfo{journal}{\emph{Proceedings of the National Academy of Sciences of the United States of America}} \bibinfo{volume}{114}, \bibinfo{number}{31} (\bibinfo{year}{2017}), \bibinfo{pages}{8148--8156}.
\newblock
\urldef\tempurl%
\url{https://doi.org/10.1073/pnas.1702996114}
\showDOI{\tempurl}
\showeprint{2017-07-13}


\bibitem[VanderWeele(2020)]%
        {vanderweele2020activities}
\bibfield{author}{\bibinfo{person}{Tyler~J VanderWeele}.} \bibinfo{year}{2020}\natexlab{}.
\newblock \showarticletitle{Activities for flourishing: An evidence-based guide}.
\newblock \bibinfo{journal}{\emph{Journal of Positive School Psychology}} \bibinfo{volume}{4}, \bibinfo{number}{1} (\bibinfo{year}{2020}), \bibinfo{pages}{79--91}.
\newblock


\bibitem[VanderWeele et~al\mbox{.}(2019)]%
        {VanderWeele2019}
\bibfield{author}{\bibinfo{person}{Tyler~J. VanderWeele}, \bibinfo{person}{Eileen McNeely}, {and} \bibinfo{person}{Howard~K. Koh}.} \bibinfo{year}{2019}\natexlab{}.
\newblock \showarticletitle{Reimagining Health-Flourishing}.
\newblock \bibinfo{journal}{\emph{JAMA}} \bibinfo{volume}{321}, \bibinfo{number}{17} (\bibinfo{year}{2019}), \bibinfo{pages}{1667--1668}.
\newblock
\urldef\tempurl%
\url{https://doi.org/10.1001/jama.2019.3035}
\showDOI{\tempurl}


\bibitem[Walsh(2011)]%
        {walsh2011lifestyle}
\bibfield{author}{\bibinfo{person}{Roger Walsh}.} \bibinfo{year}{2011}\natexlab{}.
\newblock \showarticletitle{Lifestyle and mental health.}
\newblock \bibinfo{journal}{\emph{American psychologist}} \bibinfo{volume}{66}, \bibinfo{number}{7} (\bibinfo{year}{2011}), \bibinfo{pages}{579}.
\newblock


\bibitem[Wang et~al\mbox{.}(2021)]%
        {wang2021cass}
\bibfield{author}{\bibinfo{person}{Liuping Wang}, \bibinfo{person}{Dakuo Wang}, \bibinfo{person}{Feng Tian}, \bibinfo{person}{Zhenhui Peng}, \bibinfo{person}{Xiangmin Fan}, \bibinfo{person}{Zhan Zhang}, \bibinfo{person}{Mo Yu}, \bibinfo{person}{Xiaojuan Ma}, {and} \bibinfo{person}{Hongan Wang}.} \bibinfo{year}{2021}\natexlab{}.
\newblock \showarticletitle{Cass: Towards building a social-support chatbot for online health community}.
\newblock \bibinfo{journal}{\emph{Proceedings of the ACM on Human-Computer Interaction}} \bibinfo{volume}{5}, \bibinfo{number}{CSCW1} (\bibinfo{year}{2021}), \bibinfo{pages}{1--31}.
\newblock


\bibitem[Wang et~al\mbox{.}(2015)]%
        {wang2015uncanny}
\bibfield{author}{\bibinfo{person}{Shensheng Wang}, \bibinfo{person}{Scott~O Lilienfeld}, {and} \bibinfo{person}{Philippe Rochat}.} \bibinfo{year}{2015}\natexlab{}.
\newblock \showarticletitle{The uncanny valley: Existence and explanations}.
\newblock \bibinfo{journal}{\emph{Review of General Psychology}} \bibinfo{volume}{19}, \bibinfo{number}{4} (\bibinfo{year}{2015}), \bibinfo{pages}{393--407}.
\newblock


\bibitem[Weizenbaum(1966)]%
        {weizenbaum1966eliza}
\bibfield{author}{\bibinfo{person}{Joseph Weizenbaum}.} \bibinfo{year}{1966}\natexlab{}.
\newblock \showarticletitle{ELIZA—a computer program for the study of natural language communication between man and machine}.
\newblock \bibinfo{journal}{\emph{Commun. ACM}} \bibinfo{volume}{9}, \bibinfo{number}{1} (\bibinfo{year}{1966}), \bibinfo{pages}{36--45}.
\newblock


\bibitem[Yang et~al\mbox{.}(2024)]%
        {yang2024talk2care}
\bibfield{author}{\bibinfo{person}{Ziqi Yang}, \bibinfo{person}{Xuhai Xu}, \bibinfo{person}{Bingsheng Yao}, \bibinfo{person}{Ethan Rogers}, \bibinfo{person}{Shao Zhang}, \bibinfo{person}{Stephen Intille}, \bibinfo{person}{Nawar Shara}, \bibinfo{person}{Guodong~Gordon Gao}, {and} \bibinfo{person}{Dakuo Wang}.} \bibinfo{year}{2024}\natexlab{}.
\newblock \showarticletitle{Talk2Care: An LLM-based Voice Assistant for Communication between Healthcare Providers and Older Adults}.
\newblock \bibinfo{journal}{\emph{Proceedings of the ACM on Interactive, Mobile, Wearable and Ubiquitous Technologies}} \bibinfo{volume}{8}, \bibinfo{number}{2} (\bibinfo{year}{2024}), \bibinfo{pages}{1--35}.
\newblock


\bibitem[Yao et~al\mbox{.}(2023)]%
        {yao2023development}
\bibfield{author}{\bibinfo{person}{Xuewen Yao}, \bibinfo{person}{Miriam Mikhelson}, \bibinfo{person}{S.~Craig Watkins}, \bibinfo{person}{Eunsol Choi}, \bibinfo{person}{Edison Thomaz}, {and} \bibinfo{person}{Kaya de Barbaro}.} \bibinfo{year}{2023}\natexlab{}.
\newblock \bibinfo{title}{Development and Evaluation of Three Chatbots for Postpartum Mood and Anxiety Disorders}.
\newblock
\newblock
\showeprint[arxiv]{2308.07407}~[cs.CL]


\bibitem[Yardley et~al\mbox{.}(2016)]%
        {yardley2016understanding}
\bibfield{author}{\bibinfo{person}{Lucy Yardley}, \bibinfo{person}{Bonnie~J Spring}, \bibinfo{person}{Heleen Riper}, \bibinfo{person}{Leanne~G Morrison}, \bibinfo{person}{David~H Crane}, \bibinfo{person}{Kristina Curtis}, \bibinfo{person}{Gina~C Merchant}, \bibinfo{person}{Felix Naughton}, {and} \bibinfo{person}{Ann Blandford}.} \bibinfo{year}{2016}\natexlab{}.
\newblock \showarticletitle{Understanding and promoting effective engagement with digital behavior change interventions}.
\newblock \bibinfo{journal}{\emph{American journal of preventive medicine}} \bibinfo{volume}{51}, \bibinfo{number}{5} (\bibinfo{year}{2016}), \bibinfo{pages}{833--842}.
\newblock


\bibitem[Yuan et~al\mbox{.}(2019)]%
        {yuan2019crossing}
\bibfield{author}{\bibinfo{person}{Lingyao Yuan}, \bibinfo{person}{Alan Dennis}, \bibinfo{person}{Kai Riemer}, {et~al\mbox{.}}} \bibinfo{year}{2019}\natexlab{}.
\newblock \showarticletitle{Crossing the uncanny valley? Understanding affinity, trustworthiness, and preference for more realistic virtual humans in immersive environments}.
\newblock  (\bibinfo{year}{2019}).
\newblock


\bibitem[Zhang et~al\mbox{.}(2020)]%
        {zhang2020artificial}
\bibfield{author}{\bibinfo{person}{Jingwen Zhang}, \bibinfo{person}{Yoo~Jung Oh}, \bibinfo{person}{Patrick Lange}, \bibinfo{person}{Zhou Yu}, {and} \bibinfo{person}{Yoshimi Fukuoka}.} \bibinfo{year}{2020}\natexlab{}.
\newblock \showarticletitle{Artificial intelligence chatbot behavior change model for designing artificial intelligence chatbots to promote physical activity and a healthy diet}.
\newblock \bibinfo{journal}{\emph{Journal of medical Internet research}} \bibinfo{volume}{22}, \bibinfo{number}{9} (\bibinfo{year}{2020}), \bibinfo{pages}{e22845}.
\newblock


\bibitem[Zhang(2007)]%
        {zhang2007toward}
\bibfield{author}{\bibinfo{person}{Ping Zhang}.} \bibinfo{year}{2007}\natexlab{}.
\newblock \showarticletitle{Toward a positive design theory: Principles for designing motivating information and communication technology}.
\newblock In \bibinfo{booktitle}{\emph{Designing information and organizations with a positive lens}}. Vol.~\bibinfo{volume}{2}. \bibinfo{publisher}{Emerald Group Publishing Limited}, \bibinfo{pages}{45--74}.
\newblock


\bibitem[Zhang et~al\mbox{.}(2023)]%
        {zhang-etal-2023-ask}
\bibfield{author}{\bibinfo{person}{Qiang Zhang}, \bibinfo{person}{Jason Naradowsky}, {and} \bibinfo{person}{Yusuke Miyao}.} \bibinfo{year}{2023}\natexlab{}.
\newblock \showarticletitle{Ask an Expert: Leveraging Language Models to Improve Strategic Reasoning in Goal-Oriented Dialogue Models}. In \bibinfo{booktitle}{\emph{Findings of the Association for Computational Linguistics: ACL 2023}}, \bibfield{editor}{\bibinfo{person}{Anna Rogers}, \bibinfo{person}{Jordan Boyd-Graber}, {and} \bibinfo{person}{Naoaki Okazaki}} (Eds.). \bibinfo{publisher}{Association for Computational Linguistics}, \bibinfo{address}{Toronto, Canada}, \bibinfo{pages}{6665--6694}.
\newblock
\urldef\tempurl%
\url{https://doi.org/10.18653/v1/2023.findings-acl.417}
\showDOI{\tempurl}


\bibitem[Zhang et~al\mbox{.}(2014)]%
        {zhang2014present}
\bibfield{author}{\bibinfo{person}{Ting Zhang}, \bibinfo{person}{Tami Kim}, \bibinfo{person}{Alison~Wood Brooks}, \bibinfo{person}{Francesca Gino}, {and} \bibinfo{person}{Michael~I Norton}.} \bibinfo{year}{2014}\natexlab{}.
\newblock \showarticletitle{A “present” for the future: The unexpected value of rediscovery}.
\newblock \bibinfo{journal}{\emph{Psychological science}} \bibinfo{volume}{25}, \bibinfo{number}{10} (\bibinfo{year}{2014}), \bibinfo{pages}{1851--1860}.
\newblock


\bibitem[Zhang et~al\mbox{.}(2022)]%
        {zhang2022storybuddy}
\bibfield{author}{\bibinfo{person}{Zheng Zhang}, \bibinfo{person}{Ying Xu}, \bibinfo{person}{Yanhao Wang}, \bibinfo{person}{Bingsheng Yao}, \bibinfo{person}{Daniel Ritchie}, \bibinfo{person}{Tongshuang Wu}, \bibinfo{person}{Mo Yu}, \bibinfo{person}{Dakuo Wang}, {and} \bibinfo{person}{Toby Jia-Jun Li}.} \bibinfo{year}{2022}\natexlab{}.
\newblock \showarticletitle{Storybuddy: A human-ai collaborative chatbot for parent-child interactive storytelling with flexible parental involvement}. In \bibinfo{booktitle}{\emph{Proceedings of the 2022 CHI Conference on Human Factors in Computing Systems}}. \bibinfo{pages}{1--21}.
\newblock


\bibitem[Zhao and Epley(2021)]%
        {zhao2021insufficiently}
\bibfield{author}{\bibinfo{person}{Xuan Zhao} {and} \bibinfo{person}{Nicholas Epley}.} \bibinfo{year}{2021}\natexlab{}.
\newblock \showarticletitle{Insufficiently complimentary?: Underestimating the positive impact of compliments creates a barrier to expressing them.}
\newblock \bibinfo{journal}{\emph{Journal of Personality and Social Psychology}} \bibinfo{volume}{121}, \bibinfo{number}{2} (\bibinfo{year}{2021}), \bibinfo{pages}{239}.
\newblock


\bibitem[Zhao and Epley(2022)]%
        {zhao2022surprisingly}
\bibfield{author}{\bibinfo{person}{Xuan Zhao} {and} \bibinfo{person}{Nicholas Epley}.} \bibinfo{year}{2022}\natexlab{}.
\newblock \showarticletitle{Surprisingly happy to have helped: Underestimating prosociality creates a misplaced barrier to asking for help}.
\newblock \bibinfo{journal}{\emph{Psychological Science}} \bibinfo{volume}{33}, \bibinfo{number}{10} (\bibinfo{year}{2022}), \bibinfo{pages}{1708--1731}.
\newblock


\bibitem[Zheng et~al\mbox{.}(2023)]%
        {zheng2023building}
\bibfield{author}{\bibinfo{person}{Zhonghua Zheng}, \bibinfo{person}{Lizi Liao}, \bibinfo{person}{Yang Deng}, {and} \bibinfo{person}{Liqiang Nie}.} \bibinfo{year}{2023}\natexlab{}.
\newblock \bibinfo{title}{Building Emotional Support Chatbots in the Era of LLMs}.
\newblock
\newblock
\showeprint[arxiv]{2308.11584}~[cs.CL]


\end{thebibliography}
\bibliographystyle{ACM-Reference-Format}

\clearpage
\appendix
\section*{Appendix}

\addcontentsline{toc}{section}{Appendices}

\subsection{Baseline System Interface}

\begin{figure*}[!t]
    \centering
    \subfloat[Page 1. Users log their moods.]{
    \includegraphics[width=0.98\textwidth]{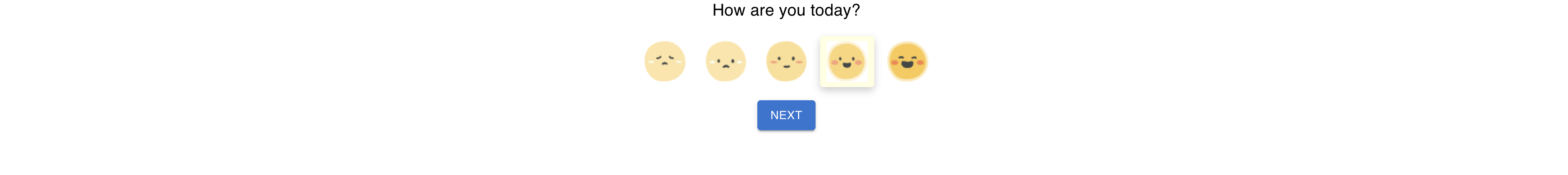}
    }
    \hfill
    \subfloat[Page 2. Select the best word(s) to describe users' feelings.]{
        \includegraphics[width=0.98\textwidth]{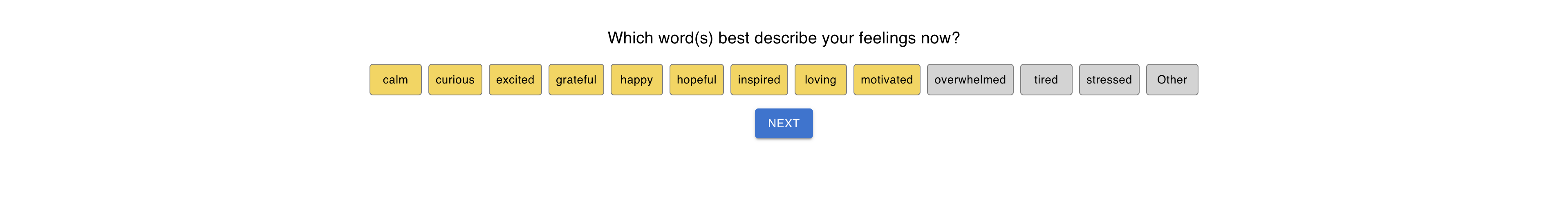}
    }
    \hfill
    \subfloat[Page 3. Input text description of their feelings.]{
        \includegraphics[width=0.98\textwidth]{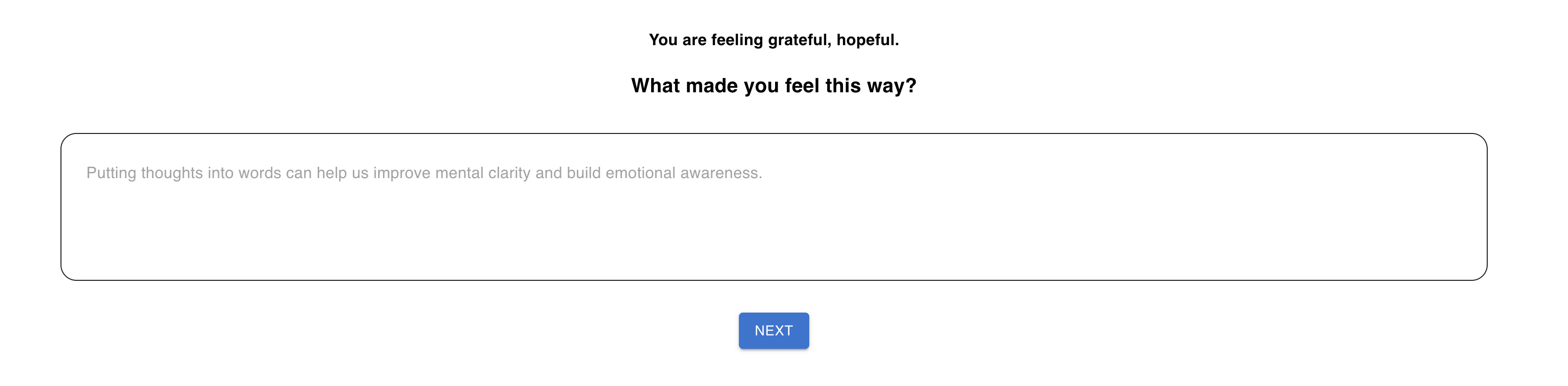}
    }


    \caption{ The user interface(s) of the baseline system. The general flow of baseline condition is identical to \system besides not having a conversation functionality and non-anthropomorphic designs. }
    
    \label{fig:app_baseline}
\end{figure*}

Figure.~\ref{fig:app_baseline} shows the interfaces we implemented for the Baseline condition. The interaction flow of the baseline condition is identical to \system besides not having a conversation functionality and non-anthropomorphic designs. After the users provide text descriptions of their feelings, they will be directed to the Typeform page of the activity recommendations, which is identical to the \system condition. 


\end{document}